\documentclass[a4paper,11pt]{article}
\pdfoutput=1

\usepackage{jcappub} 

\usepackage[T1]{fontenc} 
\usepackage{bm}
\usepackage{mathrsfs}
\usepackage{float}

\title{\boldmath Effective Field Theory for Freezing Gravity with Minimally Coupled Matter}

\author[1]{Zhibang Yao}

\affiliation[1]{Institute Lorentz, Leiden University, PO Box 9506, Leiden 2300 RA, The Netherlands}

\emailAdd{yao@lorentz.leidenuniv.nl}

\abstract{Freezing gravity (FG), a phantom-crossing dark energy model, was recently
proposed in the absence of matter. In this work, we study FG minimally
coupled to a perfect fluid through a detailed cosmological perturbation
analysis and derive the conditions for the absence of ghost and gradient
instabilities. The ghost stability conditions are obtained analytically
without taking the high-$k$ limit, yielding the ghost-free range
of scales and a finite cutoff when it exists. In the branch where
a finite cutoff is already present in vacuum, we find that matter
coupling lowers it relative to its vacuum value. Moreover, in the
high-$k$ regime, the gradient stability condition guarantees a positive
effective gravitational coupling for cold dark matter. We further
show that two characteristic properties of FG persist in the presence
of matter. First, the background and perturbation sectors remain separated,
in the sense that they are controlled by independent sets of parameters,
allowing FG to realize arbitrary background evolutions, including
phantom crossing, while maintaining stable perturbation dynamics.
Second, the scalar degree of freedom of FG becomes non-dynamical in
the large-scale limit at linear order, while propagating with a finite
speed of sound on small scales. Whether this behavior persists to
arbitrarily high orders in perturbation theory, thereby avoiding the
potential strong-coupling issue, requires further investigation through
a nonlinear perturbation analysis. To make contact with large-scale
structure and gravitational lensing observables, we also derive the
effective gravitational coupling and gravitational slip within the
quasi-static regime and express our results in terms of the EFT parameters.}

\begin{document}
\maketitle
\flushbottom

\section{Introduction \label{sec_Introduction}}

Dark energy (DE) has been a long-standing problem in cosmology, with
recent observational progress \citep{Slosar:2019flp} bringing growing
attention to it. Joint analyses of baryon acoustic oscillations (BAO)
from DESI \citep{Collaboration:2025ac,Calderon:2024aa,Lodha:2025aa}
and cosmic microwave background (CMB) data from ACT \citep{Louis:2025aa,AtacamaCosmologyTelescope:2025nti}
favor a time-varying DE component. When further combined with Type
Ia supernova (SNIa) data \citep{Collaboration:2024aa}, nonparametric
reconstructions of the equation of state (EoS) of DE show a crossing
of $w_{\text{de}}=-1$, i.e., the phantom divide, at low redshift
\citep{Lodha:2025aa,Ormondroyd:2025aa,Gu:2025aa,Akarsu:2024eoo}.
Such an implication has revived interest in studying phantom crossing
dark energy theories \citep{Orchard:2024bve,Chudaykin:2024gol,Alestas:2024gxe,Wang:2024dka,Hogas:2025ahb,Gonzalez-Fuentes:2026rgu,Calderon:2026hbr,Chanda:2026dmx,Garcia-Garcia:2026nzy,Gomez-Valent:2026ept,Ladeira:2026jne,Efstratiou:2025iqi,Liu:2025bss,Wolf:2025acj,Alam:2025epg,Pookkillath:2026wyg,Khoury:2026svx},
with the aim of identifying robust theoretical embeddings. In fact,
from a theoretical perspective, the possibility of phantom crossing
had already been explored long ago \citep{Caldwell:2002aa}, when
it was realized that quintessence, described by a single canonical
scalar field minimally coupled to matter, cannot cross the phantom
divide. It is therefore natural to explore extensions of the quintessence
framework that are capable of crossing the phantom divide, by introducing
multiple scalar fields \citep{Feng:2005aa}, non-canonical kinetic
terms \citep{Piazza2004}, non-minimal couplings \citep{Gannouji:2006aa,Amendola:2008aa,Ye:2025aa},
higher-order derivative terms \citep{Creminelli:2008wc} or other
possible modifications such as \citep{Cai:2016aa}. 

A key challenge for phantom crossing DE models is whether the underlying
theory can accommodate such behavior while maintaining dynamical stability.
As shown in \citep{Vikman:2005aa}, phantom crossing is physically
implausible in the $k$-essence framework \citep{ArmendarizPicon1999,Chiba2000},
as it either leads to gradient instability or requires an infinitely
fine-tuned trajectory in phase space. Stable phantom crossing is,
however, possible in Horndeski theory \citep{Horndeski1974,Deffayet2011},
the most general scalar-tensor theory with second-order equations
of motion, which provides a sufficiently large parameter space for
realizing such behavior \citep{Matsumoto:2017qil,Linder:2025pqt}.
On the other hand, another class of theories, known as minimally modified
gravity \citep{Lin:2017oow,DeFelice2020b}, modifies General Relativity
(GR) while propagating only two tensorial degrees of freedom (TTDOF)
\citep{Gao:2019twq,Yao2021,Yao:2023aa,Wang:2024hfd,Yu:2026fgn}. Such
theories can accommodate a broad class of cosmological backgrounds,
including phantom-crossing realizations \citep{Arora:2025msq,Scherer:2025aa,Borghetto:2026ytv,Iyonaga2021,Hiramatsu2022,Aoki2020b},
while avoiding scalar-sector instabilities associated with an additional
propagating degree of freedom.

Recently, a new DE model, dubbed freezing gravity (FG), was proposed
in \citep{Yao:2025wlx} as an alternative approach to realizing stable
phantom crossing. It is constructed within the framework of spatially
covariant gravity (SCG) theories \citep{Gao2014,Gao2019c} using the
Arnowitt-Deser-Misner (ADM) formalism. The covariant formulation of
FG can be recovered from unitary gauge by introducing a Stückelberg
field, thereby recasting it into a subclass of degenerate higher order
scalar-tensor (DHOST) theories \citep{Langlois:2015cwa,DeFelice2018}.
The main idea behind FG is the separation between the background and
perturbation sectors, in the sense that they are controlled by two
independent sets of coefficients, which allows arbitrary background
evolutions, including phantom crossing, to be realized while maintaining
stable perturbation dynamics. The realization of arbitrary background
evolutions comes from the fact that the $w_{\text{de}}$ of FG is
directly determined by two free functions of time in its Lagrangian.
Once the form of these free functions are specified, one can analyze
the phantom crossing behavior without having to first solve the background
EoM. Another characteristic feature of FG is that the sound speed
of its scalar DoF is scale-dependent and approaches infinity at large
scale, which indicates that the scalar mode becomes non-dynamical
in the infrared (IR) limit. Although a divergent speed of sound at
IR usually signals a strong coupling issue, the authors of \citep{Yao:2025wlx}
show that the scalar mode remains non-propagating up to cubic order
in the IR limit, leading them to speculate that this property persists
at any higher order. This suggests that the scalar DoF is frozen on
large scales, so that the apparent strong coupling issue is avoided.
Given this characteristic large scales behavior, the model is dubbed
“freezing gravity''. However, all the properties of FG mentioned
above were shown in vacuum, without including matter. Whether these
properties persist in the presence of matter remains an open question. 

In this work, we will address this question in FG by introducing a
minimally coupled matter sector described by a generic perfect fluid
and performing a detailed cosmological perturbation analysis of the
coupled system. For convenience in making contact with cosmological
observations, we will map FG onto the corresponding effective field
theories (EFT) of DE description \citep{Gubitosi:2012hu,Gleyzes:2014rba}
and express our results in terms of the dimensionless EFT parameters.
We will then derive the effective quadratic action for the curvature
perturbation and matter density contrast, obtain the ghost and gradient
stability conditions from which we will identify a new effective cutoff
scale corrected by the matter sector. We will also discuss whether
the large scale freezing behavior of the scalar mode survives after
the inclusion of matter sector. We will further investigate the effective
gravitational coupling \citep{Tsujikawa:2007gd} and gravitational
slip in the quasi-static regime \citep{Daniel:2008et}, providing
a direct connection to large-scale structure and gravitational lensing
observables. 

The paper is organized as follows. In section \ref{sec_Framework},
we introduce the theoretical framework of freezing gravity, the Schutz-Sorkin
action for a minimally coupled perfect fluid, and our setup for cosmological
perturbations. In section \ref{sec:Cosmological-perturbations}, we
perform the perturbation analysis to study the background evolution
and scalar perturbations. We derive the effective quadratic action
and obtain the ghost and gradient stability conditions. We also derive
the effective gravitational coupling and gravitational slip to make
contact with cosmological observables. We summarize our results and
conclude in section \ref{sec_Conclusions}.

\section{Theoretical framework \label{sec_Framework}}

In this section, we present the theoretical framework adopted in this
work. We first review the action of FG and its description in the
EFT of DE. We then introduce the Schutz-Sorkin action used to describe
a minimally coupled perfect fluid, and finally specify the setup of
perturbations employed in the subsequent cosmological perturbation
analysis. 

\subsection{Freezing gravity \label{subsec:Freezing-gravity}}

Freezing gravity (FG) was recently proposed in \citep{Yao:2025wlx}
and is described by the action 
\begin{equation}
S_{\text{fg}}=\frac{M_{\text{Pl}}^{2}}{2}\int\text{d}t\text{d}^{3}xN\sqrt{h}\Big[\sigma_{1}\left(K_{ij}K^{ij}-\frac{1}{3}K^{2}\right)+\rho_{1}R+\rho_{2}a_{i}a^{i}-\frac{2}{3}K^{2}+4\mathcal{F}K-6\mathcal{G}^{2}\big],\label{Sec-II-A_S_fg}
\end{equation}
where $R$ is the 3-dimensional Ricci scalar, while $K_{ij}$ and
$a_{i}$ denote the extrinsic curvature and 3-acceleration, respectively,
\begin{equation}
K_{ij}=\frac{1}{2N}\left(\partial_{t}h_{ij}-D_{i}N_{j}-D_{j}N_{i}\right),\quad a_{i}=D_{i}\ln N.
\end{equation}
Here $h_{ij}$, $N_{i}$ and $N$ are the induced spatial metric,
shift vector, and lapse function in the ADM formalism, with $D_{i}$
denoting the covariant derivative compatible with $h_{ij}$. The coefficients
$\sigma_{1}$, $\rho_{1}$ and $\rho_{2}$ are general functions of
time and the lapse and $\mathcal{F}$ and $\mathcal{G}$ are free
functions of time only. As we will see in the next section, the cosmological
background evolution depends only on $\mathcal{F}\left(t\right)$
and $\mathcal{G}\left(t\right)$, while the cosmological perturbations
are independently controlled by $\sigma_{1}$, $\rho_{1}$ and $\rho_{2}$.
Since the action (\ref{Sec-II-A_S_fg}) is written in the unitary
gauge, i.e. with $t=\phi$, it manifestly preserves only spatial covariance.
Its covariant formulation can be recovered by introducing the Stückelberg
field $\phi$, thereby recasting FG into a subclass of DHOST theories
in which the scalar field has a timelike profile. The corresponding
covariant formulation of FG is presented in Appendix \ref{app_FG_cov},
while we will work with the action (\ref{Sec-II-A_S_fg}) throughout
this work. 

In \citep{Yao:2025wlx}, we also derived the mapping between FG and
the EFT of DE \citep{Gubitosi:2012hu,Gleyzes:2014rba,Frusciante:2016aa,Langlois2017a}
\begin{eqnarray}
S_{\text{2}}^{\text{EFT}} & = & \int\text{d}t\text{d}^{3}xa^{3}\frac{M^{2}}{2}\Big[\delta K^{i}{}_{j}\delta K^{j}{}_{i}-\left(1+\frac{2}{3}\alpha_{\text{L}}\right)\delta K^{2}+\left(1+\alpha_{\text{H}}\right)\delta R\delta N\nonumber \\
 &  & \qquad\qquad\qquad\quad+\left(1+\alpha_{\text{T}}\right)\left(\delta_{1}R\frac{\delta\sqrt{h}}{a^{3}}+\delta_{2}R\right)+\alpha_{\text{A}}\left(\frac{\partial_{i}\delta N}{a}\right)^{2}\Big],\label{Sec-IV-A_quadrtic_action}
\end{eqnarray}
where $M$ denotes the effective Planck mass with the dimensionless
EFT parameters in $\alpha_{i}$-basis. We see that the kineticity
and braiding parameters, $\alpha_{\text{K}}$ and $\alpha_{\text{B}}$,
corresponding to the EFT of Horndeski theories \citep{Gleyzes2013},
naturally vanish in the EFT description of FG. By contrast, in the
investigations of EFT of DE \citep{Frusciante2020}, $\alpha_{\text{L}}$,
$\alpha_{\text{H}}$ and $\alpha_{\text{A}}$ are typically assumed
to be vanishing. Therefore, the study of FG can be viewed as complementary
to the usual studies of the EFT of DE. The physical interpretations
of these parameters are stated in the following.
\begin{itemize}
\item $\alpha_{\text{L}}$ characterizes the detuning between the two quadratic
extrinsic-curvature operators relative to GR. When $\alpha_{\text{L}}=0$,
their GR combination is recovered. A nonvanishing $\alpha_{\text{L}}$
commonly appears in theories with a preferred time slicing, such as
Hořava gravity \citep{Horava:2009uw} and U-DHOST theories \citep{Langlois2017a,DeFelice2018}
(see also \citep{Saito:2024aa}). 
\item $\alpha_{\text{H}}$ describes the coupling between the lapse perturbation
and the intrinsic spatial curvature. It is commonly referred to as
the beyond-Horndeski parameter, since a nonvanishing $\text{\ensuremath{\alpha_{\text{H}}}}$
implies departures from the standard Horndeski structure.
\item $\alpha_{\text{T}}$ determines the modification of the tensor propagation
speed. In the present convention, the tensor sound speed is given
by $c_{\text{T}}^{2}=1+\alpha_{\text{T}}$, so that $\alpha_{\text{T}}=0$
corresponds to luminal propagation of gravitational waves.
\item $\alpha_{\text{A}}$\footnote{It was denoted by $\beta_{3}$ in \citep{Langlois2017a,Yao:2025wlx},
whereas in this work we rename it as $\alpha_{\text{A}}$. Here the
subscript “A” refers to the 3-acceleration.} controls the spatial gradient term of the lapse perturbation coming
from the 3-acceleration $a_{i}a^{i}$. It therefore modifies the constraint
structure of the scalar sector.
\end{itemize}
The mapping between the dimensionless EFT parameters and the coefficients
of the FG action (\ref{Sec-II-A_S_fg}) is given by 
\begin{equation}
M^{2}=M_{\text{Pl}}^{2}\bar{\sigma}_{1},\quad\alpha_{\text{L}}=\frac{1}{\bar{\sigma}_{1}}-1,\quad\alpha_{\text{T}}=\frac{\bar{\rho}_{1}}{\bar{\sigma}_{1}}-1,\quad\alpha_{\text{H}}=\frac{\bar{\rho}_{1}+\partial_{N}\bar{\rho}_{1}}{\bar{\sigma}_{1}}-1,\quad\alpha_{\text{A}}=\frac{\bar{\rho}_{2}}{\bar{\sigma}_{1}},\label{Sec-IV-A_EFT_params}
\end{equation}
where the bars here and hereafter indicate that the corresponding
quantities are evaluated on the cosmological background. In particular,
we denote $\partial_{N}\bar{\rho}_{1}\equiv\frac{\partial\rho_{1}}{\partial N}\left[t,\bar{N}\left(t\right)\right]$.
For convenience in making contact with cosmological phenomenology,
we will instead express our results in terms of the EFT parameters
throughout this work. The corresponding expressions in terms of the
original action coefficients can be straightforwardly recovered from
the mapping above. From eq. (\ref{Sec-IV-A_EFT_params}), we see that
$M^{2}$ and $\alpha_{\text{L}}$ are related by $\bar{\sigma}_{1}$
and are therefore not independent in FG, satisfying 
\begin{equation}
M^{2}\left(1+\alpha_{\text{L}}\right)=M_{\text{Pl}}^{2},\label{Sec-II-A_M_a_L}
\end{equation}
whereas they are considered as independent functions in the EFT formalism.
In order to keep our results as general as possible, we will treat
$M^{2}$ and $\alpha_{\text{L}}$ as independent quantities throughout
the derivations and impose the FG relation (\ref{Sec-II-A_M_a_L})
only when stated explicitly. In the previous work \citep{Yao:2025wlx},
FG was constructed in the absence of matter. In the next subsection,
we introduce a matter sector minimally coupled to it.

\subsection{Schutz-Sorkin action\label{subsec:Schutz-Sorkin-action}}

For the matter sector, we consider a perfect fluid minimally coupled
to the FG action (\ref{Sec-II-A_S_fg}). We employ the Schutz-Sorkin
action \citep{Schutz:1977df} (see also \citep{Brown:1993aa,Felice:2010aa,Felice:2016aa,Aoki:2025aa})
as a convenient variational formulation of a relativistic perfect
fluid, without the need to specify its fundamental origin. The Schutz-Sorkin
action is written as 
\begin{equation}
S_{\text{m}}=-\int\text{d}^{4}x\left[\sqrt{-g}\rho_{\text{m}}\left(n_{\text{m}}\right)+J^{\mu}\nabla_{\mu}\ell\right],\label{II-B_Schutz-Sorkin}
\end{equation}
where $\rho_{\text{m}}$ denotes the matter energy density as a function
of its particle number density 
\begin{equation}
n_{\text{m}}=\sqrt{\frac{g_{\mu\nu}J^{\mu}J^{\nu}}{g}},\label{II-B_n_m}
\end{equation}
and $J^{\mu}$ is the particle-number flux density, while $\ell$
is a Lagrange multiplier. By varying the Schutz-Sorkin action (\ref{II-B_Schutz-Sorkin})
with respect to $g^{\mu\nu}$, $\ell$ and $J^{\mu}$, one obtains
\begin{equation}
T_{\mu\nu}^{\left(\text{m}\right)}=\frac{-2}{\sqrt{-g}}\frac{\delta S_{\text{m}}}{\delta g^{\mu\nu}}=\left(\rho_{\text{m}}+p_{\text{m}}\right)\left(u_{\text{m}}\right)_{\mu}\left(u_{\text{m}}\right)_{\nu}+p_{\text{m}}g_{\mu\nu},\label{II-B_T_munu}
\end{equation}
and
\begin{equation}
\nabla_{\mu}J^{\mu}=0,\quad\nabla_{\mu}\ell=\left(u_{\text{m}}\right)_{\mu}\rho_{\text{m},n_{\text{m}}},\label{II-B_nablaJ}
\end{equation}
where the pressure and normalized 4-velocity of the perfect fluid
are defined, respectively, by 

\begin{equation}
p_{\text{m}}=n_{\text{m}}\rho_{\text{m},n_{\text{m}}}-\rho_{\text{m}},\quad\left(u_{\text{m}}\right)^{\mu}=\frac{J^{\mu}}{n_{\text{m}}\sqrt{-g}},\label{II-B_4-velo}
\end{equation}
with $\rho_{\text{m},n_{\text{m}}}\equiv\text{d}\rho_{\text{m}}/\text{d}n_{\text{m}}$.
We see that the action (\ref{II-B_Schutz-Sorkin}) indeed gives rise
to the energy-momentum tensor $T_{\mu\nu}^{\left(\text{m}\right)}$
of a perfect fluid and its flux density $J^{\mu}$ is conserved. The
Lagrange multiplier $\ell$ is determined by eq. (\ref{II-B_nablaJ}),
which will be used when performing cosmological perturbations in the
matter sector. The combination of the FG action (\ref{Sec-II-A_S_fg})
and the Schutz-Sorkin action (\ref{II-B_Schutz-Sorkin}) defines the
total action considered in this work, i.e.,
\begin{equation}
S_{\text{tot}}=S_{\text{fg}}+S_{\text{m}}.\label{II-B_S_tot}
\end{equation}
Before performing the cosmological perturbation analysis of the total
action (\ref{II-B_S_tot}), we briefly clarify our perturbation setup
in the next subsection.

\subsection{Perturbations setup \label{Subsec:pertur_setup}}

In this subsection, we specify our setup for the perturbation variables
and gauge choice. For the purposes of this work, we focus on scalar
perturbations and neglect the vector and tensor perturbations. Therefore,
we parametrize the ADM variables as
\begin{equation}
N=\bar{N}e^{A},\quad N_{i}=\bar{N}a\partial_{i}B,\quad h_{ij}=a^{2}e^{2\zeta}\delta_{ij},\label{II-A_N=00003DNA}
\end{equation}
where $\bar{N}\left(t\right)$ denotes the background lapse, $a\left(t\right)$
is the scale factor, and $\left\{ A,B,\zeta\right\} $ are the scalar
perturbations. Note that we have set the scalar shear potential $E=0$
as a gauge choice. We also point out that, in general, $\bar{N}\left(t\right)$
cannot be chosen freely since we are working in unitary gauge, hence
time reparametrization symmetry is fixed. However, as shown in \citep{Yao:2025wlx},
this symmetry is restored on the cosmological background in FG, so
that the value of $\bar{N}\left(t\right)$ can be again chosen freely.
We will confirm this property in the next section; for the moment,
we keep $\bar{N}$ as a general function of time.

For the matter sector, the flux vector $J^{\mu}$ and Lagrange multiplier
$\ell$ are formally perturbed as 
\begin{equation}
J^{0}=\bar{J}^{0}+\delta J^{0},\quad J^{i}=\bar{J}^{i}+\delta J^{i},\quad\ell=\bar{\ell}+\delta\ell.\label{II-C_J=00003DJb+Jp}
\end{equation}
In order to translate the formal quantities into physical variables,
we perform the following calculations.
\begin{itemize}
\item First, using the definition (\ref{II-B_n_m}) gives the number density
perturbation at linear order as
\begin{equation}
\bar{n}_{\text{m}}+\delta n_{\text{m}}=\frac{\bar{J}^{0}}{a^{3}}\left(1-3\zeta+\frac{\delta J^{0}}{\bar{J}^{0}}\right),
\end{equation}
from which we identify 
\begin{equation}
\bar{J}^{0}=\bar{n}_{\text{m}}a^{3}\equiv\mathcal{N}_{0},\quad\delta J^{0}=3\mathcal{N}_{0}\zeta+\frac{a^{3}}{\bar{\rho}_{\text{m},n_{\text{m}}}}\delta\rho_{\text{m}},\label{II-C_J0}
\end{equation}
where $\mathcal{N}_{0}$ denotes the conserved background particle
number and in the second relation we have used 
\begin{equation}
\delta n_{\text{m}}=\frac{\delta\rho_{\text{m}}}{\bar{\rho}_{\text{m},n_{\text{m}}}}.
\end{equation}
\item Second, background isotropy requires $\bar{J}^{i}=0$, while the scalar
part of $\delta J^{i}$ can be written as
\begin{equation}
\delta J^{i}=\frac{\delta^{ik}\partial_{k}\delta j}{a^{2}},
\end{equation}
where $\delta j$ denotes the scalar perturbations associated with
$J^{i}$. Using the linear perturbation of the 4-velocity in eq. (\ref{II-B_4-velo})
gives 
\begin{equation}
\left(\delta u_{\text{m}}\right)_{i}=a\partial_{i}B+\left(\bar{N}\mathcal{N}_{0}\right)^{-1}\partial_{i}\delta j\equiv-\partial_{i}v_{m},
\end{equation}
where we introduce the velocity potential $v_{m}$ such that 
\begin{equation}
\delta j=-\bar{N}\mathcal{N}_{0}\left(v_{m}+aB\right).\label{II-C_dj}
\end{equation}
\item Lastly, integrating eq. (\ref{II-B_nablaJ}) gives
\begin{equation}
\bar{\ell}=-\int_{t_{0}}^{t}\text{d}t\bar{N}\bar{\rho}_{\text{m},n_{\text{m}}},\quad\delta\ell=-\bar{\rho}_{\text{m},n_{\text{m}}}v_{\text{m}},\label{II-C_l}
\end{equation}
where the integration constant has been chosen accordingly. 
\end{itemize}
To summarize, the perturbations of flux vector, $\delta J^{0}$ and
$\delta J^{i}$, have been expressed in terms of the density perturbation
$\delta\rho_{\text{m}}$ and the velocity potential $v_{m}$ respectively,
while $\bar{J}^{i}$, $\bar{\ell}$ and $\delta\ell$ have been determined
accordingly. The background quantity $\bar{J}^{0}$, which is identical
to the conserved background particle number $\mathcal{N}_{0}$, does
not appear in the final result, since it only corresponds to the normalization
of the number of background particles. For later convenience, we further
define the EoS parameter, sound speed, and density contrast of matter
sector by
\begin{equation}
w_{\text{m}}\equiv\frac{\bar{p}_{\text{m}}}{\bar{\rho}_{\text{m}}},\quad c_{\text{m}}^{2}\equiv\frac{\bar{n}_{\text{m}}\bar{\rho}_{\text{m},n_{\text{m}}n_{\text{m}}}}{\bar{\rho}_{\text{m},n_{\text{m}}}},\quad\delta_{\text{m}}\equiv\frac{\delta\rho_{\text{m}}}{\bar{\rho}_{\text{m}}}.\label{II-C_wm}
\end{equation}
As a result, the scalar perturbations of the full action (\ref{II-B_S_tot})
are described by $\left\{ A,B,\zeta,\delta_{\text{m}},v_{m}\right\} $,
with coefficients depending on the EFT parameters (\ref{Sec-IV-A_EFT_params})
and $\left\{ H,\bar{\rho}_{\text{m}},w_{\text{m}},c_{\text{m}}^{2}\right\} $.
Now we are ready to perform a detailed cosmological perturbation analysis
of FG minimally coupled to a matter sector in the next section.

\section{Cosmological perturbations\label{sec:Cosmological-perturbations}}

In this section, we will perform a detailed cosmological perturbation
analysis of the total action (\ref{II-B_S_tot}) with the variables
introduced above. In particular, we will study the background evolution
of the model, the linear stability conditions of scalar perturbations
and the effective gravitational coupling as well as the gravitational
slip in the quasi-static regime.

\subsection{Background evolution\label{subsec:Background-evolution}}

On the FLRW background, the total action (\ref{II-B_S_tot}) reduces
to
\begin{equation}
S_{\text{tot}}^{\left(0\right)}=-\int\text{d}t\text{d}^{3}x\bar{N}a^{3}\left[3M_{\text{Pl}}^{2}\left(H^{2}-2H\mathcal{F}+\mathcal{G}^{2}\right)-\bar{p}_{\text{m}}\right],\label{III-A_S_BG}
\end{equation}
where the Hubble parameter is defined by $H\equiv\frac{1}{\bar{N}a}\frac{\text{d}a}{\text{d}t}$
and $\bar{p}_{\text{m}}$ is the background value of matter pressure
defined in eq. (\ref{II-B_4-velo}). We see that, even in the presence
of matter, only $\mathcal{F}\left(t\right)$ and $\mathcal{G}\left(t\right)$
in the FG action (\ref{Sec-II-A_S_fg}), enter the background dynamics
and we will show below, are associated with the energy density and
pressure of the DE component. Another property evident from the action
(\ref{III-A_S_BG}) is that time reparametrization symmetry is restored
on the FLRW background in FG. This follows from the fact that both
the Hubble parameter and the integration measure are invariant under
time reparametrization, rendering the entire background action (\ref{III-A_S_BG})
invariant as well. Thanks to this property, we are allowed to set
$\bar{N}=1$ henceforth.

The background EoM are obtained by varying the background action (\ref{III-A_S_BG})
with respect to $\bar{N}$, $a$ and $\bar{n}_{\text{m}}$, respectively,
\begin{eqnarray}
3H^{2} & = & M_{\text{Pl}}^{-2}\left(\bar{\rho}_{\text{de}}+\bar{\rho}_{\text{m}}\right),\label{III-A_BG_eom_1}\\
-3H^{2}-2\dot{H} & = & M_{\text{Pl}}^{-2}\left(\bar{p}_{\text{de}}+\bar{p}_{\text{m}}\right),\label{III-A_BG_eom_2}\\
\dot{\bar{\rho}}_{\text{m}} & = & -3H\left(\bar{\rho}_{\text{m}}+\bar{p}_{\text{m}}\right),\label{III-A_BG_eom_3}
\end{eqnarray}
where we define 
\begin{equation}
\bar{\rho}_{\text{de}}\equiv3M_{\text{Pl}}^{2}\mathcal{G}^{2},\quad\bar{p}_{\text{de}}\equiv-M_{\text{Pl}}^{2}\left(3\mathcal{G}^{2}+2\dot{\mathcal{F}}\right).\label{III-A_rho_de_p_de}
\end{equation}
We see that eq. (\ref{III-A_BG_eom_3}) is nothing but the standard
continuity equation for the matter sector, and eq. (\ref{III-A_BG_eom_1})
and (\ref{III-A_BG_eom_2}) take the usual form of the Friedmann equations
with the effective energy density and pressure of DE component identified
in eq. (\ref{III-A_rho_de_p_de}). We can further define the EoS of
DE by 
\begin{equation}
w_{\text{de}}=\frac{\bar{p}_{\text{de}}}{\bar{\rho}_{\text{de}}}=-\frac{2}{3}\frac{\dot{\mathcal{F}}}{\mathcal{G}^{2}}-1,
\end{equation}
which is determined solely by the two free functions of time, $\mathcal{F}\left(t\right)$
and $\mathcal{G}\left(t\right)$ appearing in the action (\ref{Sec-II-A_S_fg}).
Consequently, once these two functions are specified, the evolution
of the EoS can be obtained directly, without the need to solve the
background EoM. Furthermore, whether the EoS of DE crosses the phantom
divide, $w_{\text{de}}=-1$, can be determined simply by checking
whether $\dot{\mathcal{F}}$ changes sign. 

Next we derive the quadratic action for scalar perturbations for the
total action (\ref{II-B_S_tot}). Before doing this, we point out
that the free functions, $\mathcal{F}\left(t\right)$ and $\mathcal{G}\left(t\right)$,
can be determined from the Friedmann equations (\ref{III-A_BG_eom_1})
and (\ref{III-A_BG_eom_2}), and do not enter the perturbation dynamics.\footnote{Another example in which a general solution to its Friedmann equations
is obtainable was also recently discussed in \citep{Hell:2025lgn}.} In this sense, the background evolution and perturbations in FG are
independently controlled by two distinct sets of parameters, realizing
a separation between the background and perturbation sectors. This
background perturbation separation provides FG with more flexibility
in fitting observational data than conventional models. 

\subsection{Effective quadratic action\label{subsec:Effective-quadratic-action} }

Based on the perturbation setup introduced in subsection \ref{Subsec:pertur_setup}
for our cosmological perturbations, we derive the quadratic action
for scalar perturbations of the total action (\ref{II-B_S_tot}) as
follows
\begin{eqnarray}
S_{\text{scalar}}^{\left(2\right)} & = & \int\text{d}t\frac{\text{d}^{3}k}{\left(2\pi\right)^{3}}a^{3}M^{2}\Big\{(\alpha_{\text{L}}+1)\left[\left(6\dot{\zeta}-3HA+2\frac{\mathit{k}^{2}}{a^{2}}aB\right)HA-\left(3\dot{\zeta}+2\frac{\mathit{k}^{2}}{a^{2}}aB\right)\dot{\zeta}\right]\nonumber \\
 &  & +\frac{\mathit{k}^{2}}{a^{2}}\left[\frac{1}{2}\alpha_{\text{A}}A^{2}+2\left(1+\alpha_{\text{H}}\right)A\zeta+\left(1+\alpha_{\text{T}}\right)\zeta^{2}-2\alpha_{\text{L}}k^{2}B^{2}\right]-\frac{\bar{\rho}_{\text{m}}}{M^{2}}\Big[\dot{\delta}_{\text{m}}v_{\text{m}}+(w_{\text{m}}+1)\nonumber \\
 &  & \times\left(3\dot{\zeta}+\frac{\mathit{k}^{2}}{a^{2}}aB+\frac{1}{2}\frac{\mathit{k}^{2}}{a^{2}}v_{\text{m}}\right)v_{\text{m}}+\left(A+\frac{c_{\text{m}}^{2}\delta_{\text{m}}}{2(w_{\text{m}}+1)}+\left(c_{\text{m}}^{2}-w_{\text{m}}\right)3Hv_{\text{m}}\right)\delta_{\text{m}}\Big]\Big\},\label{III_B_S_scalar}
\end{eqnarray}
where we have transformed to the Fourier space and adopted the EFT
parameters defined in eq. (\ref{Sec-IV-A_EFT_params}). As expected,
the quadratic action (\ref{III_B_S_scalar}) is controlled by the
EFT parameters $\left\{ M^{2},\alpha_{\text{L}},\alpha_{\text{H}},\alpha_{\text{T}},\alpha_{\text{A}}\right\} $,
while the free functions $\mathcal{F}\left(t\right)$ and $\mathcal{G}\left(t\right)$
do not appear. We also see that $\left\{ A,B,v_{\text{m}}\right\} $
play the role of auxiliary fields in the quadratic action (\ref{III_B_S_scalar}),
and their EoM yield the following constraint equations 
\begin{eqnarray}
6(\alpha_{\text{L}}+1)(\dot{\zeta}-AH)+2\alpha_{\text{L}}\frac{\mathit{k}^{2}}{a^{2}}aB+3(w_{\text{m}}+1)\frac{\bar{\rho}_{\text{m}}}{M^{2}}v_{\text{m}} & = & 0,\label{III_B_eom_A}\\
6H(\alpha_{\text{L}}+1)\left(\dot{\zeta}-AH+\frac{1}{3}\frac{\mathit{k}^{2}}{a^{2}}aB\right)+\frac{\mathit{k}^{2}}{a^{2}}\left[\alpha_{\text{A}}A+2(\alpha_{\text{H}}+1)\zeta\right]-\frac{\bar{\rho}_{\text{m}}}{M^{2}}\delta_{\text{m}} & = & 0,\label{III_B_eom_B}\\
\dot{\delta}_{\text{m}}+\left(c_{\text{m}}^{2}-w_{\text{m}}\right)3H\delta_{\text{m}}+\left(w_{\text{m}}+1\right)\left(3\dot{\zeta}+\frac{\mathit{k}^{2}}{a^{2}}aB+\frac{\mathit{k}^{2}}{a^{2}}v_{\text{m}}\right) & = & 0.\label{III_B_eom_vm}
\end{eqnarray}
After solving for the auxiliary fields from eq. (\ref{III_B_eom_A})-(\ref{III_B_eom_vm})
and substituting their solutions back into the action (\ref{III_B_S_scalar}),
we derive the effective quadratic action for the curvature perturbation
and matter density contrast as follows
\begin{equation}
S_{\text{eff}}^{\left(2\right)}=\int\text{d}t\frac{\text{d}^{3}k}{\left(2\pi\right)^{3}}a^{3}\left(\dot{\chi}_{\vec{k}}^{I}\mathcal{K}_{IJ}\dot{\chi}_{-\vec{k}}^{J}-\frac{k^{2}}{a^{2}}\chi_{\vec{k}}^{I}\mathcal{G}_{IJ}\chi_{-\vec{k}}^{J}-\frac{k}{a}\dot{\chi}_{\vec{k}}^{I}\mathcal{I}_{IJ}\chi_{-\vec{k}}^{J}\right),\label{III-B_S2_eff}
\end{equation}
where we introduce a unified notation for the dynamical fields
\begin{equation}
\chi_{\vec{k}}^{I}\equiv\left(\begin{array}{c}
\zeta\\
\frac{\delta_{\text{m}}}{k}
\end{array}\right)\left(t,\vec{k}\right),\quad I=\zeta,\delta,
\end{equation}
and rescale $\delta_{\text{m}}$ by $k$ such that the diagonal components
of the kinetic matrix $\mathcal{K}_{IJ}$ exhibit the canonical behavior
$\mathcal{O}\left(k^{0}\right)$ in the high-$k$ limit. By definition,
the kinetic and gradient matrices $\mathcal{K}_{IJ}$ and $\mathcal{G}_{IJ}$
are symmetric, while the mixed matrix $\mathcal{I}_{IJ}$ is taken
to be antisymmetric via integration by parts. The nonvanishing components
of these matrices are listed below.
\begin{eqnarray}
\mathcal{K}_{\zeta\zeta} & \equiv & \Xi^{-1}3M^{2}\Big[18H^{2}(\alpha_{\text{L}}+1)(w_{\text{m}}+1)\bar{\rho}_{\text{m}}\nonumber \\
 &  & -3(w_{\text{m}}+1)\bar{\rho}_{\text{m}}\alpha_{\text{A}}\frac{k^{2}}{a^{2}}+2M^{2}(\alpha_{\text{L}}+1)\alpha_{\text{A}}\frac{k^{4}}{a^{4}}\Big],\label{III-B_K_zz}\\
\mathcal{K}_{\delta\delta} & \equiv & \frac{M^{2}\bar{\rho}_{\text{m}}}{\Xi\left(w_{\text{m}}+1\right)}\left[6H^{2}(\alpha_{\text{L}}+1)+\alpha_{\text{L}}\alpha_{\text{A}}\frac{k^{2}}{a^{2}}\right]\frac{k^{2}}{a^{2}},\label{III-B_K_dd}\\
\mathcal{K}_{\zeta\delta} & \equiv & \Xi^{-1}3M^{2}\bar{\rho}_{\text{m}}\left[6H^{2}(\alpha_{\text{L}}+1)-\alpha_{\text{A}}\frac{k^{2}}{a^{2}}\right]\frac{k}{a},\label{III-B_K_zd}\\
\mathcal{I}_{\zeta\delta} & \equiv & \Xi^{-1}3aHM^{2}\bar{\rho}_{\text{m}}\Big\{18H^{2}(\alpha_{\text{L}}+1)\left(c_{\text{m}}^{2}-w_{\text{m}}\right)\nonumber \\
 &  & +\left[2\alpha_{\text{H}}(\alpha_{\text{L}}+1)-3\alpha_{\text{A}}\left(c_{\text{m}}^{2}-w_{\text{m}}\right)\right]\frac{k^{2}}{a^{2}}\Big\},\label{III-B_I_zd}
\end{eqnarray}
and
\begin{equation}
\mathcal{G}_{\zeta\zeta}\equiv\frac{G_{\zeta\zeta}^{\left(0\right)}+G_{\zeta\zeta}^{\left(2\right)}\frac{k^{2}}{a^{2}}+G_{\zeta\zeta}^{\left(4\right)}\frac{k^{4}}{a^{4}}+G_{\zeta\zeta}^{\left(6\right)}\frac{k^{6}}{a^{6}}+G_{\zeta\zeta}^{\left(8\right)}\frac{k^{8}}{a^{8}}}{\Xi^{2}},\label{III-B_G_zz}
\end{equation}
\begin{equation}
\mathcal{G}_{\delta\delta}\equiv\frac{G_{\delta\delta}^{\left(0\right)}+G_{\delta\delta}^{\left(2\right)}\frac{k^{2}}{a^{2}}+G_{\delta\delta}^{\left(4\right)}\frac{k^{4}}{a^{4}}+G_{\delta\delta}^{\left(6\right)}\frac{k^{6}}{a^{6}}+G_{\delta\delta}^{\left(8\right)}\frac{k^{8}}{a^{8}}}{2M^{2}\left(1+w_{\text{m}}\right)\Xi^{2}},\label{III-B_G_dd}
\end{equation}
\begin{equation}
\mathcal{G}_{\zeta\delta}\equiv\frac{G_{\zeta\delta}^{\left(0\right)}+G_{\zeta\delta}^{\left(2\right)}\frac{k^{2}}{a^{2}}+G_{\zeta\delta}^{\left(4\right)}\frac{k^{4}}{a^{4}}+G_{\zeta\delta}^{\left(6\right)}\frac{k^{6}}{a^{6}}+G_{\zeta\delta}^{\left(8\right)}\frac{k^{8}}{a^{8}}}{2\frac{k}{a}\Xi^{2}},\label{III-B_G_zd}
\end{equation}
with 
\begin{eqnarray}
\Xi & \equiv & 18H^{2}(\alpha_{\text{L}}+1)(w_{\text{m}}+1)\bar{\rho}_{\text{m}}+2M^{2}\alpha_{\text{L}}\alpha_{\text{A}}\frac{k^{4}}{a^{4}}\nonumber \\
 &  & +3\left[4H^{2}M^{2}(\alpha_{\text{L}}+1)-\alpha_{\text{A}}(w_{\text{m}}+1)\bar{\rho}_{\text{m}}\right]\frac{k^{2}}{a^{2}}.\label{III-B_Xi}
\end{eqnarray}
We point out that more general effective quadratic actions with additional
EFT parameters are derived in \citep{Felice:2016aa,Frusciante:2016aa},
which include our result as a special case.

At this stage, we have obtained the effective quadratic action (\ref{III-B_S2_eff})
for the scalar perturbations of FG minimally coupled to matter. We
see that the gravity and matter sectors are nontrivially coupled through
the off-diagonal components of the kinetic, gradient, and mixed matrices,
which makes the stability analysis more involved than in the vacuum
case. Nevertheless, the ghost and gradient stability conditions can
still be systematically derived from the effective quadratic action
(\ref{III-B_S2_eff}), as we will show in the following subsections. 

\subsection{Ghost stability conditions\label{subsec:No-ghost-conditions}}

The absence of ghosts requires the kinetic matrix $\mathcal{K}_{IJ}$
in the effective quadratic action (\ref{III-B_S2_eff}) to be positive
definite. In the present case, $\mathcal{K}_{IJ}$ is a 2\texttimes 2
symmetric matrix, its positive definiteness is equivalent to requiring
\begin{equation}
\mathcal{K}_{\delta\delta}>0,\quad\text{det}\mathcal{K}>0,\label{III-C_K_dd_det_K}
\end{equation}
where $\text{det}\mathcal{K}$ denotes the determinant of the kinetic
matrix $\mathcal{K}_{IJ}$. Using eq. (\ref{III-B_K_zz})-(\ref{III-B_K_zd}),
they can be written as
\begin{equation}
\mathcal{K}_{\delta\delta}=\frac{M^{2}\bar{\rho}_{\text{m}}\mathcal{Q}}{\Xi\left(1+w_{\text{m}}\right)}\frac{k^{2}}{a^{2}},\qquad\text{det}\mathcal{K}_{IJ}=\frac{3a^{2}M^{4}\bar{\rho}_{\text{m}}\alpha_{\text{A}}\left(1+\alpha_{\text{L}}\right)}{\Xi\left(1+w_{\text{m}}\right)}\frac{k^{4}}{a^{4}},\label{III-C_KQ_detK}
\end{equation}
where we have defined
\begin{equation}
\mathcal{Q}\equiv6H^{2}\left(1+\alpha_{\text{L}}\right)+\alpha_{\text{L}}\alpha_{\text{A}}\frac{k^{2}}{a^{2}}.\label{III-C_Q}
\end{equation}

Assuming $M^{2}>0$, $\bar{\rho}_{\text{m}}>0$, and $w_{\text{m}}>-1$,
which are reasonable conditions for a physically viable gravity and
ordinary matter sector, the freezing-gravity relation (\ref{Sec-II-A_M_a_L})
implies 
\begin{equation}
\alpha_{\text{L}}>-1.\label{III-C_alpha_L>-1}
\end{equation}
Furthermore, according to eq. (\ref{III-B_Xi}), $\Xi>0$ in the low-$k$
limit. Therefore, for sufficiently small but nonzero $k$, the sign
of $\text{det}\mathcal{K}_{IJ}$ is determined by $\alpha_{\text{A}}$.
The no-ghost condition $\text{det}\mathcal{K}_{IJ}>0$ thus requires
\begin{equation}
\alpha_{\text{A}}>0.
\end{equation}
With $\alpha_{\text{L}}>-1$ and $\alpha_{\text{A}}>0$, the no-ghost
conditions (\ref{III-C_K_dd_det_K}) simply reduce to
\begin{equation}
\mathcal{Q}\left(k\right)>0,\quad\Xi\left(k\right)>0,\label{III-C_Q=000026Xi}
\end{equation}
where $\mathcal{Q}\left(k\right)$ and $\Xi\left(k\right)$ are linear
and quadratic functions of $k^{2}/a^{2}$, respectively. The ghost
instability problem therefore reduces to determining the range of
$k$ over which both inequalities are simultaneously satisfied.
\begin{itemize}
\item For $-1<\alpha_{\text{L}}<0$, solving 
\begin{equation}
\mathcal{Q}\left(k_{\text{b}}\right)=0,\quad\Xi\left(k_{\text{c}}\right)=0,
\end{equation}
yields two different scales 
\begin{equation}
\frac{k_{\text{b}}^{2}}{a^{2}}=-\frac{6H^{2}\left(1+\alpha_{\text{L}}\right)}{\alpha_{\text{L}}\alpha_{\text{A}}},\quad\frac{k_{\text{c}}^{2}}{a^{2}}=\frac{k_{\text{b}}^{2}}{a^{2}}\frac{1}{2}\left[1-\epsilon_{\text{m}}+\sqrt{\left(1-\epsilon_{\text{m}}\right)^{2}-4\alpha_{\text{L}}\epsilon_{\text{m}}}\right],\label{III-C_k_b_k_c}
\end{equation}
where we denote
\begin{equation}
\epsilon_{\text{m}}\equiv\frac{\alpha_{\text{A}}\left(1+w_{\text{m}}\right)\bar{\rho}_{\text{m}}}{4M^{2}H^{2}\left(1+\alpha_{\text{L}}\right)}.
\end{equation}
The scale $k_{\text{b}}$ coincides with the cutoff found for FG in
vacuum \citep{Yao:2025wlx}, and we therefore refer to it as the ``bare''
cutoff. In contrast, $k_{\text{c}}$ incorporates the correction from
the matter sector and will be referred to as the ``corrected'' cutoff.
Since $-1<\alpha_{\text{L}}<0$ and $\epsilon_{\text{m}}>0$, we have
\begin{equation}
\sqrt{\left(1-\epsilon_{\text{m}}\right)^{2}-4\alpha_{\text{L}}\epsilon_{\text{m}}}<1+\epsilon_{\text{m}},
\end{equation}
which immediately implies 
\begin{equation}
k_{\text{c}}<k_{\text{b}}.
\end{equation}
In particular, in the vacuum limit $\bar{\rho}_{\text{m}}\rightarrow0$,
one has $\epsilon_{\text{m}}\rightarrow0$ and hence $k_{\text{c}}\rightarrow k_{\text{b}}$.
Thus, the minimal coupling to matter lowers the effective cutoff of
FG relative to its vacuum value. Since $k_{\text{c}}<k_{\text{b}}$,
$\mathcal{Q}\left(k\right)>0$ throughout the whole regime $0\le k<k_{\text{c}}$
and thus imposes no additional restriction on the ghost-free region.
\item For $\alpha_{\text{L}}>0$, $\mathcal{Q}\left(k\right)>0$ is automatically
satisfied for all $k$, so that the ghost-stability region is determined
solely by $\Xi\left(k\right)>0$. In this case, a finite cutoff $k_{\text{c}}$,
determined by $\Xi\left(k_{\text{c}}\right)=0$, exists when
\begin{equation}
\epsilon_{\text{m}}\ge\epsilon_{\text{m}}^{+}\quad\text{ with}\quad\epsilon_{\text{m}}^{+}\equiv\left(\sqrt{1+\alpha_{\text{L}}}+\sqrt{\alpha_{\text{L}}}\right)^{2}.\label{III-C_e_m+}
\end{equation}
Otherwise, if $\epsilon_{\text{m}}<\epsilon_{\text{m}}^{+}$, the
theory remains ghost-free for all $k$ within the regime of validity
of the EFT. 
\item For the marginal case, $\alpha_{\text{L}}=0$, one has $\mathcal{Q}=6H^{2}>0$
and $\epsilon_{\text{m}}^{+}=1$, while $\Xi\left(k\right)$ reduces
to a linear function of $k^{2}/a^{2}$. A finite cutoff exists only
for $\epsilon_{\text{m}}>\epsilon_{\text{m}}^{+}$, whereas for $\epsilon_{\text{m}}\le\epsilon_{\text{m}}^{+}$
the theory remains ghost-free for all $k$.
\end{itemize}
We summarize that, under $\alpha_{\text{L}}>-1$ and $\alpha_{\text{A}}>0$,
the no-ghost conditions (\ref{III-C_K_dd_det_K}) require eq. (\ref{III-C_Q=000026Xi}).
Depending on the parameter region, these conditions either define
a finite cutoff scale $k_{\text{c}}$, below which the theory remains
ghost-free, or allow the theory to remain ghost-free throughout the
regime of validity of the EFT.

Before closing this subsection, we discuss the dynamics of scalar
mode in the FG here. It was shown in \citep{Yao:2025wlx} that, in
the vacuum case, the kinetic term of $\zeta$ vanishes as $k\rightarrow0$,
indicating that the scalar mode becomes non-propagating in the large-scale
limit. Although such behavior may in general signal a strong-coupling
issue, the higher-order analysis performed there showed that the scalar
mode remains non-propagating in the IR regime, suggesting that it
is frozen on large scales. It is therefore important to investigate
whether this property persists after coupling with matter. According
to eq. (\ref{III-C_KQ_detK}) with (\ref{III-B_Xi}), in the $k\rightarrow0$
limit, we find

\begin{equation}
\lim_{k\rightarrow0}\det\mathcal{K}=\lim_{k\rightarrow0}\frac{M^{4}a^{2}\alpha_{\text{A}}}{6H^{2}\left(1+w_{\text{m}}\right)^{2}}\frac{k^{4}}{a^{4}}\rightarrow0,\label{III-C_detK->0}
\end{equation}
showing that the kinetic matrix is degenerate in the large-scale limit.\footnote{One may wonder whether the $k^{4}/a^{4}$ dependence can be absorbed
by a field redefinition. However, in this way, the redefined field
itself is diluted as the large-scale limit is approached.} Since the matter sector is assumed to possess regular dynamics, this
degeneracy can be attributed to the gravitational sector. This indicates
that, at least at linear order, the scalar mode of FG remains non-propagating
on large scales even in the presence of minimally coupled matter.
Whether this freezing property persists at nonlinear orders requires
a higher-order perturbative analysis or a full Hamiltonian analysis,
which we leave for future work. We next turn to the gradient stability
conditions for the effective quadratic action (\ref{III-B_S2_eff}).

\subsection{Gradient stability conditions\label{subsec:Gradient-stability-conditions}}

To investigate gradient stability, we consider the $k\rightarrow\infty$
limit, since gradient instabilities become increasingly severe toward
small scales and the propagation speeds of the scalar modes are locally
well defined in the short-wavelength regime. According to the analysis
in the previous subsection, the branch with $-1<\alpha_{\text{L}}<0$
necessarily encounters a finite cutoff $k_{\text{c}}$, so that the
$k\rightarrow\infty$ limit lies beyond the regime of validity of
the EFT. By contrast, for $\alpha_{\text{L}}>0$ with $\epsilon_{\text{m}}<\epsilon_{\text{m}}^{+}$,
the theory can remain ghost-free up to arbitrarily large $k$. We
therefore restrict $\alpha_{\text{L}}>0$ in the following analysis
of gradient stability, while the $-1<\alpha_{\text{L}}<0$ branch
will be briefly discussed at the end of this subsection.

In the high-$k$ limit, the matrix elements in eq. (\ref{III-B_K_zz})-(\ref{III-B_G_zd})
admit the following $k$-scaling: 
\begin{eqnarray}
\mathcal{K}_{\zeta\zeta} & \sim & \mathcal{K}_{\zeta\zeta}^{\left(0\right)}+\mathcal{O}\left(\mathit{k}^{-2}\right),\quad\mathcal{K}_{\delta\delta}\sim\mathcal{K}_{\delta\delta}^{\left(0\right)}+\mathcal{O}\left(\mathit{k}^{-2}\right),\quad\mathcal{K}_{\zeta\delta}\sim\mathcal{K}_{\zeta\delta}^{\left(-1\right)}k^{-1}+\mathcal{O}\left(k^{-3}\right),\nonumber \\
\mathcal{G}_{\zeta\zeta} & \sim & \mathcal{G}_{\zeta\zeta}^{\left(0\right)}+\mathcal{G}_{\zeta\zeta}^{\left(-2\right)}\mathit{k}^{-2}+\mathcal{O}\left(k^{-4}\right),\quad\mathcal{G}_{\delta\delta}\sim\mathcal{G}_{\delta\delta}^{\left(0\right)}+\mathcal{G}_{\delta\delta}^{\left(-2\right)}\mathit{k}^{-2}+\mathcal{O}\left(k^{-4}\right),\nonumber \\
\mathcal{G}_{\zeta\delta} & \sim & \mathcal{G}_{\zeta\delta}^{\left(-1\right)}k^{-1}+\mathcal{O}\left(k^{-3}\right),\quad\mathcal{I}_{\zeta\delta}\sim\mathcal{I}_{\zeta\delta}^{\left(-2\right)}k^{-2}+\mathcal{O}\left(k^{-4}\right).\label{III-D_I_zd}
\end{eqnarray}
Note that we assume $\alpha_{\text{L}}\alpha_{\text{A}}\neq0$ in
deriving the high-$k$ expansions above. If either parameter vanishes,
the leading $k$-scaling behavior may change, and the corresponding
branch should be analyzed separately (see \citep{Felice:2016aa,Frusciante:2016aa}
for more details). In the present case, the leading order coefficients
are 
\begin{equation}
\mathcal{K}_{\delta\delta}^{\left(0\right)}=\frac{a^{2}\bar{\rho}_{\text{m}}}{2\left(1+w_{\text{m}}\right)},\quad\mathcal{G}_{\delta\delta}^{\left(0\right)}=\frac{a^{2}\bar{\rho}_{\text{m}}c_{\text{m}}^{2}}{2\left(1+w_{\text{m}}\right)},\label{III-D_K_dd_G_dd}
\end{equation}
and
\begin{equation}
\mathcal{K}_{\zeta\zeta}^{\left(0\right)}=\frac{3M^{2}\left(1+\alpha_{\text{L}}\right)}{\alpha_{\text{L}}},\quad\mathcal{G}_{\zeta\zeta}^{\left(0\right)}=M^{2}\left[\frac{2}{\alpha_{\text{A}}}\left(1+\alpha_{\text{H}}\right)^{2}-\left(1+\alpha_{\text{T}}\right)\right].\label{III-D_K_zz_G_zz}
\end{equation}
The remaining expansion coefficients are collected in Appendix \ref{App_series_coeff}
due to their length. We retain the next-to-leading terms $\mathcal{G}_{\zeta\zeta}^{\left(-2\right)}$
and $\mathcal{G}_{\delta\delta}^{\left(-2\right)}$, since they contribute
to the effective mass of $\zeta$ and $\delta_{\text{m}}$. From eq.
(\ref{III-D_I_zd}), we see that, at leading order in the high-$k$
regime, the effective quadratic action (\ref{III-B_S2_eff}) separates
into decoupled gravity and matter sectors, because all the off-diagonal
matrix elements are suppressed by inverse powers of $k$. 

The corresponding sound speeds can therefore be defined locally as
\begin{equation}
c_{\text{s}}^{2}\equiv\frac{\mathcal{G}_{\zeta\zeta}^{\left(0\right)}}{\mathcal{K}_{\zeta\zeta}^{\left(0\right)}},\quad c_{\text{m}}^{2}\equiv\frac{\mathcal{G}_{\delta\delta}^{\left(0\right)}}{\mathcal{K}_{\delta\delta}^{\left(0\right)}}.\label{III-D_c^2}
\end{equation}
For matter sector, the second relation trivially reproduces the definition
of $c_{\text{m}}^{2}$ in eq. (\ref{II-C_wm}), and the absence of
gradient instability requires $c_{\text{m}}^{2}>0.$ For gravity sector,
the speed of sound is given by 
\begin{equation}
c_{\text{s}}^{2}=\frac{\alpha_{\text{L}}\left[2\left(1+\alpha_{\text{H}}\right)^{2}-\alpha_{\text{A}}\left(1+\alpha_{\text{T}}\right)\right]}{3\alpha_{\text{A}}\left(1+\alpha_{\text{L}}\right)},\label{III-D_c_s^2}
\end{equation}
where $\alpha_{\text{L}}>0$ and $\alpha_{\text{A}}>0$ have been
imposed in the preceding analysis. Requiring $c_{\text{s}}^{2}>0$
then yields the gradient stability condition in high-$k$ limit as
follows 
\begin{equation}
\left(1+\alpha_{\text{H}}\right)^{2}>\frac{\alpha_{\text{A}}}{2}\left(1+\alpha_{\text{T}}\right)>0,\label{III-D_gradient_cond}
\end{equation}
where the second inequality follows from the gradient stability condition
for tensor perturbations, $c_{\text{T}}^{2}=1+\alpha_{\text{T}}>0$. 

Before the end of this subsection, we briefly discuss the gradient
stability of the other branch, $-1<\alpha_{\text{L}}<0$. In this
case, instead of taking the high-$k$ limit, one may formally investigate
the propagation of the scalar perturbations within the Wentzel--Kramers--Brillouin
(WKB) approximation, provided that the modes remain below the cutoff
and satisfy the WKB condition. Applying the WKB approximation to the
quadratic action (\ref{III-B_S2_eff}), the generalized dispersion
relation can be written as

\begin{equation}
\text{det}\left[-\omega^{2}\mathcal{K}_{IJ}+\text{i}\omega\frac{k}{a}\mathcal{I}_{IJ}+\frac{k^{2}}{a^{2}}\mathcal{G}_{IJ}\right]=0,\label{III-D_dispersion}
\end{equation}
where $\omega$ denotes the angular frequency and is assumed to satisfy
$\left|\omega\right|\gg H$ within WKB regime. Defining $\omega^{2}=c_{\pm}^{2}k^{2}/a^{2}$,
the two propagation eigenvalues are

\begin{equation}
c_{\pm}^{2}=\frac{\mathcal{D}\pm\sqrt{\mathcal{D}^{2}-\text{det}\mathcal{K}\text{det}\mathcal{G}}}{\text{det}\mathcal{K}},\label{III-D_c_s^2_pm}
\end{equation}
where $\text{det}\mathcal{G}$ denotes the determinant of the gradient
matrix $\mathcal{G}_{IJ}$ and
\begin{equation}
\mathcal{D}\equiv\frac{1}{2}\left(\mathcal{K}_{\zeta\zeta}\mathcal{G}_{\delta\delta}+\mathcal{K}_{\delta\delta}\mathcal{G}_{\zeta\zeta}-2\mathcal{K}_{\zeta\delta}\mathcal{G}_{\zeta\delta}+\mathcal{I}_{\zeta\delta}^{2}\right).
\end{equation}
Here we have used the facts that $\mathcal{K}_{IJ}$ and $\mathcal{G}_{IJ}$
are symmetric, whereas $\mathcal{I}_{IJ}$ is anti-symmetric. Therefore,
within the WKB regime, the absence of gradient instabilities requires

\begin{equation}
\mathcal{D}^{2}-\text{det}\mathcal{K}\text{det}\mathcal{G}\ge0,\quad c_{\pm}^{2}>0.\label{III-D_stability_WKB}
\end{equation}
This result is also applicable to the branch $\alpha_{\text{L}}>0$.
In fact, one can verify that taking the high-$k$ limit to eq. (\ref{III-D_stability_WKB})
reproduces the gradient stability condition eq. (\ref{III-D_gradient_cond})
with $c_{\text{m}}^{2}>0$.

We emphasize that the sound speed derived above characterizes the
local propagation of the scalar mode in the short-wavelength regime,
$k/a\gg H$, and therefore does not contradict the freezing behavior
found in the large-scale limit. The scalar mode of FG can thus be
non-propagating on large scales while remaining dynamical on sufficiently
small scales. This scale-dependent behavior also suggests that FG
can leave nontrivial signatures toward small cosmological scales.
To characterize such effects, in the next subsections we investigate
the effective gravitational coupling and the gravitational slip parameter.

\subsection{Effective gravitational coupling\label{subsec:Effective-gravitational-coupling}}

In this subsection, we derive the effective gravitational coupling
\citep{Tsujikawa:2007gd} relevant to the growth of large scale structure
within the quasi-static approximation (QSA). We consider the regime
\citep{Sawicki2015}
\begin{equation}
\frac{H}{c_{\text{s}}}\ll\frac{k}{a}\lesssim\frac{H}{c_{\text{m}}},
\end{equation}
while remaining below the cutoff $k_{\text{c}}$, when present. The
lower bound ensures that the curvature perturbation is well inside
its sound horizon, so that the time-derivative terms associated with
the slowly evolving solution are suppressed relative to the spatial-gradient
terms, whereas the upper bound keeps the matter perturbation slowly
evolving. The existence of such a regime requires $c_{\text{m}}^{2}$
to be sufficiently small compared with $c_{\text{s}}^{2}$, as naturally
realized for cold dark matter (CDM), for which $c_{\text{m}}=0$.
For generality, we nevertheless keep the matter sector in the form
of a generic perfect fluid throughout the derivation and specialize
to CDM only when identifying the effective gravitational coupling
relevant to structure growth.

Varying the effective quadratic action (\ref{III-B_S2_eff}) w.r.t.
$\zeta$ and $\delta_{\text{m}}$ yields the equations of motion 
\begin{eqnarray}
0 & = & \ddot{\zeta}+\left(3H+\frac{\dot{\mathcal{K}}_{\zeta\zeta}}{\mathcal{K}_{\zeta\zeta}}\right)\dot{\zeta}+\frac{k^{2}}{a^{2}}\frac{\mathcal{G}_{\zeta\zeta}}{\mathcal{K}_{\zeta\zeta}}\zeta+\frac{\mathcal{K}_{\delta\zeta}}{\mathcal{K}_{\zeta\zeta}}\frac{\ddot{\delta}_{\text{m}}}{k}\nonumber \\
 &  & +\left(\frac{3H\mathcal{K}_{\delta\zeta}+\mathcal{\dot{K}}_{\delta\zeta}}{\mathcal{K}_{\zeta\zeta}}+\frac{k}{a}\frac{\mathcal{I}_{\delta\zeta}}{\mathcal{K}_{\zeta\zeta}}\right)\frac{\dot{\delta}_{\text{m}}}{k}+\left(\frac{k^{2}}{a^{2}}\frac{\mathcal{G}_{\delta\zeta}}{\mathcal{K}_{\zeta\zeta}}+\frac{k}{a}\frac{H\mathcal{I}_{\delta\zeta}+\frac{1}{2}\mathcal{\dot{I}}_{\delta\zeta}}{\mathcal{K}_{\zeta\zeta}}\right)\frac{\delta_{\text{m}}}{k},\label{III-D_EoM_zeta}
\end{eqnarray}
and
\begin{eqnarray}
0 & = & \ddot{\delta}_{\text{m}}+\left(3H+\frac{\dot{\mathcal{K}}_{\delta\delta}}{\mathcal{K}_{\delta\delta}}\right)\dot{\delta}_{\text{m}}+\frac{k^{2}}{a^{2}}\frac{\mathcal{G}_{\delta\delta}}{\mathcal{K}_{\delta\delta}}\delta_{\text{m}}+\frac{\mathcal{K}_{\zeta\delta}}{\mathcal{K}_{\delta\delta}}k\ddot{\zeta}\nonumber \\
 &  & +\left(\frac{2H\mathcal{K}_{\zeta\delta}+\dot{\mathcal{K}}_{\zeta\delta}}{\mathcal{K}_{\delta\delta}}+\frac{k}{a}\frac{\mathcal{I}_{\zeta\delta}}{\mathcal{K}_{\delta\delta}}\right)k\dot{\zeta}+\left(\frac{k^{2}}{a^{2}}\frac{\mathcal{G}_{\zeta\delta}}{\mathcal{K}_{\delta\delta}}+\frac{k}{a}\frac{H\mathcal{I}_{\zeta\delta}+\frac{1}{2}\dot{\mathcal{I}}_{\zeta\delta}}{\mathcal{K}_{\delta\delta}}\right)k\zeta,\label{III-D_EoM_delta}
\end{eqnarray}
where the coefficients are given by eq. (\ref{III-B_K_zz}) - (\ref{III-B_Xi}).
In the QSA regime, keeping the leading contributions according to
the high-$k$ expansion, eq. (\ref{III-D_EoM_zeta}) and (\ref{III-D_EoM_delta})
reduce to 

\begin{equation}
\frac{k^{2}}{a^{2}}\frac{\mathcal{G}_{\zeta\zeta}^{\left(0\right)}}{\mathcal{K}_{\zeta\zeta}^{\left(0\right)}}\zeta+\frac{k^{2}}{a^{2}}\frac{\mathcal{G}_{\delta\zeta}^{\left(-1\right)}k^{-1}}{\mathcal{K}_{\zeta\zeta}^{\left(0\right)}}\frac{\delta_{\text{m}}}{k}=0,\label{zeta=00003Ddelta/k}
\end{equation}
and 
\begin{equation}
\ddot{\delta}_{\text{m}}+\left(3H+\frac{\dot{\mathcal{K}}_{\delta\delta}^{\left(0\right)}}{\mathcal{K}_{\delta\delta}^{\left(0\right)}}\right)\dot{\delta}_{\text{m}}+\frac{k^{2}}{a^{2}}\frac{\mathcal{G}_{\delta\delta}^{\left(0\right)}+\mathcal{G}_{\delta\delta}^{\left(-2\right)}k^{-2}}{\mathcal{K}_{\delta\delta}^{\left(0\right)}}\delta_{\text{m}}+\frac{k^{2}}{a^{2}}\frac{\mathcal{G}_{\zeta\delta}^{\left(-1\right)}k^{-1}}{\mathcal{K}_{\delta\delta}^{\left(0\right)}}k\zeta=0,\label{III-D_ddot_delta_m}
\end{equation}
where the series coefficients are defined in eq. (\ref{III-D_I_zd}),
with their explicit expressions collected in Appendix \ref{App_series_coeff}. 

From eq. (\ref{zeta=00003Ddelta/k}), we obtain
\begin{equation}
\zeta=-\frac{a^{2}\left(1+\alpha_{\text{H}}\right)\bar{\rho}_{\text{m}}\delta_{\text{m}}}{k^{2}M^{2}\left[2\left(1+\alpha_{\text{H}}\right)^{2}-\alpha_{\text{A}}\left(1+\alpha_{\text{T}}\right)\right]},\label{III-D_zeta=00003Ddelta}
\end{equation}
and we see that, within the QSA, the curvature perturbation is sourced
by the matter density contrast through a Poisson-like constraint.
Substituting eq. (\ref{zeta=00003Ddelta/k}) into (\ref{III-D_ddot_delta_m})
yields
\begin{equation}
\ddot{\delta}_{\text{m}}+\mathcal{Y}\dot{\delta}_{\text{m}}-4\pi G_{\text{eff}}\bar{\rho}_{\text{m}}\delta_{\text{m}}=0,\label{III-D_ddot=00007Bdelta=00007D_m}
\end{equation}
where the effective friction coefficient $\mathcal{Y}$ and effective
gravitational coupling $G_{\text{eff}}$ are defined as 
\begin{equation}
\mathcal{Y}\equiv3H+\left.\left(\frac{\dot{\mathcal{K}}_{\delta\delta}}{\mathcal{K}_{\delta\delta}}\right)\right|^{\left(0\right)},\quad G_{\text{eff}}\equiv-\frac{1}{4\pi\bar{\rho}_{\text{m}}}\left.\left(\frac{k^{2}}{a^{2}}\frac{\text{det}\mathcal{G}}{\mathcal{K}_{\delta\delta}\mathcal{G}_{\zeta\zeta}}\right)\right|^{\left(0\right)}.\label{III-D_Y=000026G_eff}
\end{equation}
Here ``$\left.\right|^{\left(0\right)}$'' denotes the leading contribution
in the high-$k$ expansion. Up to this point, neither the FG relation
(\ref{Sec-II-A_M_a_L}) nor the CDM specialization has been imposed.
Therefore, eq. (\ref{III-D_Y=000026G_eff}) also applies to the EFT
action (\ref{Sec-IV-A_quadrtic_action}) coupled to a generic perfect
fluid, provided the relevant modes lie within the QSA regime.

To obtain the effective gravitational coupling relevant to structure
growth, we now specialize to CDM by setting $w_{\text{m}}=0$ and
$c_{\text{m}}=0$. In this case, eq. (\ref{III-D_Y=000026G_eff})
becomes
\begin{equation}
\mathcal{Y}_{\text{cdm}}=2H,\quad G_{\text{cdm}}=\left(8\pi M^{2}\right)^{-1}\left[\frac{\left(1+\alpha_{\text{H}}\right)^{2}}{1+\alpha_{\text{T}}}-\frac{\alpha_{\text{A}}}{2}\right]^{-1},\label{III-D_G_cdm}
\end{equation}
from which we see that the effective friction term retains its standard
form, while the modification of structure growth is entirely encoded
in $G_{\text{cdm}}$ at leading order in the QSA. As shown in \citep{Aoki:2025aa},
nontrivial interactions between the DE and DM sectors can induce a
modification of the effective friction term. We also notice that the
gradient stability condition derived in eq. (\ref{III-D_gradient_cond})
implies $G_{\text{cdm}}>0$. Therefore, the absence of gradient instabilities
guarantees $G_{\text{cdm}}>0$, ensuring an attractive effective gravitational
interaction for CDM in FG.

\subsection{Gravitational slip\label{subsec:Gravitational-slip}}

While the effective gravitational coupling characterizes the strength
of gravity relevant to the growth of matter perturbations, the gravitational
slip describes the relative response of the two scalar metric potentials.
It therefore provides complementary information on modifications of
gravity, particularly when combining observables associated with large
scale structure and gravitational lensing. 

In Newtonian gauge, the gravitational slip is defined by
\begin{equation}
\eta\equiv\frac{\text{\ensuremath{\Psi}}}{\Phi},\label{III-F_eta}
\end{equation}
where $\Psi$ and $\Phi$ denote the temporal and spatial Bardeen
potentials, respectively. To express $\eta$ in terms of the EFT parameters,
we use the gauge transformation between Newtonian and unitary gauges
\begin{equation}
\Psi=A+\left(aB\right)^{\cdot},\quad\Phi=-\zeta-aHB,\label{III-F_gauge_transf}
\end{equation}
where $A$, $B$ and $\zeta$ are defined in eq. (\ref{II-A_N=00003DNA}).
We could use the constraint equations (\ref{III_B_eom_A})-(\ref{III_B_eom_vm})
for solving the auxiliary fields $A$ and $B$ as follows 

\begin{eqnarray}
A & = & \mathcal{E}_{A\dot{\zeta}}\dot{\zeta}+\mathcal{E}_{A\zeta}\zeta+\mathcal{E}_{A\dot{\delta}}\dot{\delta}_{\text{m}}+\mathcal{E}_{A\delta}\delta_{\text{m}},\label{III-F_eom_A}\\
B & = & \mathcal{E}_{B\dot{\zeta}}\dot{\zeta}+\mathcal{E}_{B\zeta}\zeta+\mathcal{E}_{B\dot{\delta}}\dot{\delta}_{\text{m}}+\mathcal{E}_{B\delta}\delta_{\text{m}},\label{III-F_eom_B}
\end{eqnarray}
where the coefficients are given in Appendix \ref{App_constr_eq}.
To obtain the gravitational slip relevant to cosmological observables,
we first consider the subhorizon regime and expand eq. (\ref{III-F_eom_A})
and (\ref{III-F_eom_B}) in the high-$k$ limit as

\begin{equation}
A\sim\mathcal{E}_{A\zeta}^{\left(0\right)}\zeta+\mathcal{E}_{A\delta}^{\left(-2\right)}k^{-2}\delta_{\text{m}},\quad B\sim\mathcal{O}\left(k^{-2}\right),\label{III-F_A-B}
\end{equation}
which shows that $B$ is suppressed at leading order, while $A$ receives
contributions proportional to $\zeta$ and $\delta_{\text{m}}/k^{2}$
with 
\begin{equation}
\mathcal{E}_{A\zeta}^{\left(0\right)}\equiv-\frac{2\left(1+\alpha_{\text{H}}\right)}{\alpha_{\text{A}}},\quad\mathcal{E}_{A\delta}^{\left(-2\right)}\equiv\frac{a^{2}\bar{\rho}_{\text{m}}}{M^{2}\alpha_{\text{A}}}.\label{III-F_EAzeta}
\end{equation}
We then impose the QSA relation (\ref{III-D_zeta=00003Ddelta}) and
substitute eq. (\ref{III-F_gauge_transf}), (\ref{III-F_A-B}) and
(\ref{III-F_EAzeta}) into (\ref{III-F_eta}). As a result, we obtain
\begin{equation}
\eta=\frac{1+\alpha_{\text{T}}}{1+\alpha_{\text{H}}},\label{III-E_slip}
\end{equation}
from which we see that the gravitational slip in FG is only determined
by $\alpha_{\text{H}}$ and $\alpha_{\text{T}}$.

Combining the effective gravitational coupling (\ref{III-D_G_cdm})
with the gravitational slip (\ref{III-E_slip}) allows us to identify
the phenomenological functions $\mu$ and $\Sigma$, defined by \citep{Pogosian:2016pwr}
\begin{eqnarray}
\frac{k^{2}}{a^{2}}\Psi & = & -4\pi G_{\text{N}}\mu\bar{\rho}_{\text{m}}\Delta_{\text{m}},\\
\frac{k^{2}}{a^{2}}\left(\Psi+\Phi\right) & = & -8\pi G_{\text{N}}\Sigma\bar{\rho}_{\text{m}}\Delta_{\text{m}},
\end{eqnarray}
where $G_{\text{N}}=\left(8\pi M_{\text{Pl}}^{2}\right)^{-1}$ is
the Newtonian gravitational constant and $\Delta_{\text{m}}$ is the
comoving density contrast, which agrees with $\delta_{\text{m}}$
at the leading order in subhorizon regime. With our convention $\eta=\Psi/\Phi$,
these quantities are related to $G_{\text{eff}}$ and $\eta$ through 

\begin{equation}
\mu=\frac{G_{\text{eff}}}{G_{\text{N}}},\quad\Sigma=\frac{\mu}{2}\left(1+\frac{1}{\eta}\right).
\end{equation}
Therefore, in FG, we derive
\begin{eqnarray}
\mu & = & \frac{M_{\text{Pl}}^{2}}{M^{2}}\left[\frac{\left(1+\alpha_{\text{H}}\right)^{2}}{1+\alpha_{\text{T}}}-\frac{\alpha_{\text{A}}}{2}\right]^{-1},\\
\Sigma & = & \frac{M_{\text{Pl}}^{2}}{M^{2}}\left[\frac{2+\alpha_{\text{H}}+\alpha_{\text{T}}}{2\left(1+\alpha_{\text{H}}\right)^{2}-\alpha_{\text{A}}\left(1+\alpha_{\text{T}}\right)}\right],
\end{eqnarray}
where we have used eq. (\ref{III-D_G_cdm}) and (\ref{III-E_slip}).
Up to this point, we have established the main dynamical and phenomenological
properties of freezing gravity in the presence of matter. We now summarize
our results and conclude.

\section{Conclusions \label{sec_Conclusions}}

In this work, we have performed a detailed cosmological perturbation
analysis of freezing gravity (FG) defined by (\ref{Sec-II-A_S_fg}),
a phantom crossing dark energy (DE) model proposed in \citep{Yao:2025wlx},
in the presence of a minimally coupled matter sector. We derive the
conditions for the absence of ghost and gradient instabilities and
identify an effective cutoff scale from the ghost stability conditions.
In the quasi-static regime, we further find that the modification
of structure growth is encoded in an effective gravitational coupling,
which is guaranteed to be positive for cold dark matter by the gradient
stability condition, while a gravitational slip can arise from nonluminal
gravitational-wave propagation and/or a beyond-Horndeski structure.

On the cosmological background, we show that the inclusion of minimally
coupled matter does not spoil the characteristic separation between
the background and perturbation sectors of FG. The background evolution
is still controlled solely by the two free functions $\mathcal{F}\left(t\right)$
and $\mathcal{G}\left(t\right)$, which determine the effective density,
pressure, and equation of state (EoS) of the DE sector, while the
perturbation dynamics are governed independently by the EFT parameters
(\ref{Sec-IV-A_EFT_params}) and the physical quantities of the matter
sector. This allows us to identify phantom crossing without having
to solve the background equations of motion, while providing FG with
more flexibility in fitting cosmological data.

In sections \ref{subsec:No-ghost-conditions} and \ref{subsec:Gradient-stability-conditions},
we discussed the ghost and gradient instabilities of FG in the presence
of matter. The corresponding ghost and gradient stability conditions
are formally given in eq. (\ref{III-C_Q=000026Xi}) and (\ref{III-D_stability_WKB}),
respectively. The no-ghost conditions (\ref{III-C_Q=000026Xi}) are
obtained analytically in closed form and determine the ghost-free
range of scales. Depending on the parameter region, they either define
a finite cutoff scale or allow the theory to remain ghost-free throughout
the regime of validity of the EFT. In particular, for $-1<\alpha_{\text{L}}<0$,
the matter-corrected cutoff $k_{\text{c}}$ is lower than its vacuum
value $k_{\text{b}}$. The gradient stability conditions (\ref{III-D_stability_WKB})
are obtained within the WKB approximation and apply below the cutoff
whenever a finite cutoff is present. In particular, for the branch
$\alpha_{\text{L}}>0$, the high-$k$ limit can be consistently taken,
in which case the conditions (\ref{III-D_stability_WKB}) reduce to
eq. (\ref{III-D_gradient_cond}). 

In sections \ref{subsec:Effective-gravitational-coupling} and \ref{subsec:Gravitational-slip},
we further investigated the effective gravitational coupling and gravitational
slip in the quasi-static regime. For a generic perfect fluid, we first
derived the formal expressions for the effective friction coefficient
and the effective gravitational coupling in eq. (\ref{III-D_Y=000026G_eff}).
We then specialized to CDM and found that the effective friction term
retains its standard form, while the modification of structure growth
is entirely encoded in $G_{\text{cdm}}$, as shown in eq. (\ref{III-D_G_cdm}).
Moreover, we found that the gradient stability condition (\ref{III-D_gradient_cond})
guarantees $G_{\text{cdm}}>0$, ensuring an attractive effective gravitational
interaction. In eq. (\ref{III-E_slip}), we also obtained the gravitational
slip parameter $\eta$, which depends only on $\alpha_{\text{H}}$
and $\alpha_{\text{T}}$. We also translate these results into the
$\mu$-$\Sigma$ formalism, providing a direct connection to large
scale structure and gravitational lensing observables.

We conclude with several remarks. First, as a characteristic of FG,
the scalar DoF was speculated in our previous work \citep{Yao:2025wlx}
to become frozen in the large scale limit in vacuum. In eq. (\ref{III-C_detK->0}),
we show that this behavior persists at linear order even after the
inclusion of matter, as indicated by the degeneracy of the kinetic
matrix in the IR limit. Whether this property persists at arbitrarily
higher order in cosmological perturbation, thereby avoiding the strong-coupling
issue, remains an open question. A robust demonstration would require
a nonlinear perturbation analysis or a full Hamiltonian analysis.
Second, it will be important to confront FG with cosmological observations,
in particular with the DESI BAO data. In fact, FG has already been
implemented in the Einstein--Boltzmann solver $\mathtt{EFTCAMB}$
\citep{Hu:2014aa,Raveri:2014aa,Ye:2026qqf}, providing a direct route
to a systematic comparison with current and future cosmological data.
Such an analysis will allow us to constrain the EFT parameters of
FG, identify the preferred region of parameter space, and investigate
possible signatures of the model in the matter and CMB power spectra.
It will also be interesting to study how the effective gravitational
coupling and gravitational slip derived in this work affect the growth
of large scale structure and how they can be constrained by observational
data. Finally, the independent control of the background and perturbation
sectors in FG may also provide a useful framework for addressing other
cosmological tensions, such as the Hubble tension \citep{Cai:2026swf}
and the $S_{8}$ tension \citep{Pantos:2026koc}. Since FG is not
restricted to low redshifts, its phenomenology may extend to earlier
cosmological epochs. We leave these for future work. 

\appendix

\section{The covariant action of freezing gravity \label{app_FG_cov}}

Freezing gravity (\ref{Sec-II-A_S_fg}) can be recast into a class
of scalar-tensor theories of the form (see \citep{Yao:2025wlx} for
more details)

\begin{equation}
S_{\text{fg}}^{\left(\text{cov}\right)}=\int\text{d}^{4}x\sqrt{-g}\left[G\left(\phi,X\right)\mathscr{R}+P\left(\phi,X\right)+\mathcal{B}^{\mu\nu}\phi_{\mu\nu}+\mathcal{C}^{\mu\nu\rho\sigma}\phi_{\mu\nu}\phi_{\rho\sigma}\right],\label{Sec-II_S_U-DHOST}
\end{equation}
where $\mathscr{R}$ denotes the 4-dimensional Ricci scalar and 
\begin{equation}
\mathcal{B}^{\mu\nu}\equiv B_{1}g^{\mu\nu}+B_{2}\phi^{\mu}\phi^{\nu},\label{Sec-II_B^munu}
\end{equation}
and
\begin{eqnarray}
\mathcal{C}^{\mu\nu\rho\sigma} & \equiv & \frac{1}{2}C_{1}\left(g^{\mu\rho}g^{\nu\sigma}+g^{\mu\sigma}g^{\nu\rho}\right)+C_{2}g^{\mu\nu}g^{\rho\sigma}+\frac{1}{2}C_{3}\left(\phi^{\mu}\phi^{\nu}g^{\rho\sigma}+\phi^{\rho}\phi^{\sigma}g^{\mu\nu}\right)\nonumber \\
 &  & +\frac{1}{4}C_{4}\left(\phi^{\mu}\phi^{\rho}g^{\nu\sigma}+\phi^{\nu}\phi^{\rho}g^{\mu\sigma}+\phi^{\mu}\phi^{\sigma}g^{\nu\rho}+\phi^{\nu}\phi^{\sigma}g^{\mu\rho}\right)+C_{5}\phi^{\mu}\phi^{\nu}\phi^{\rho}\phi^{\sigma}.
\end{eqnarray}
Here the coefficients $\left\{ G,P,B_{1},B_{2},C_{1,2,3,4,5}\right\} $
are the general functions of scalar field $\phi$ and its kinetic
term $X\equiv-\frac{1}{2}\phi_{\mu}\phi^{\mu}$. We adopt the notations
$\phi_{\mu}\equiv\nabla_{\mu}\phi$, $\phi_{\mu\nu}\equiv\nabla_{\nu}\nabla_{\mu}\phi$,
etc. The relations among the coefficients are given as follows 

\begin{equation}
P=-4\sqrt{2X}\mathcal{F}_{\phi}-6\mathcal{G}^{2},\quad B_{1}=-2G{}_{\phi},\quad B_{2}=-\frac{1}{X}G{}_{\phi},
\end{equation}
\begin{equation}
C_{2}=\frac{1}{3X}\left(G-XC_{1}-1\right),\quad C_{3}=\frac{1}{2X^{2}}\left(2XC_{2}+\frac{1}{\sqrt{2X}}G{}_{X}\right),
\end{equation}
\begin{equation}
C_{5}=\frac{1}{12X^{3}}\left(6X^{2}C_{4}-4XC_{1}+\frac{3}{\sqrt{2X}}G{}_{X}+G-1\right),
\end{equation}
The maps between the coefficients of the actions (\ref{Sec-II-A_S_fg})
and (\ref{Sec-II_S_U-DHOST}) are also given here

\begin{equation}
P=-4\sqrt{2X}\mathcal{F}_{\phi}-6\mathcal{G}^{2},\quad G=\rho_{1},\quad B_{1}=-2\rho_{1}{}_{\phi},\quad B_{2}=-\frac{1}{X}\rho_{1}{}_{\phi},
\end{equation}
\begin{equation}
C_{1}=\frac{1}{2X}\left(\sigma_{1}-\rho_{1}\right),\quad C_{2}=\frac{1}{3X}\left(\rho_{1}-XC_{1}-1\right),\quad C_{5}=\frac{1}{12X^{3}}\left(\sigma_{1}-1+\frac{3}{2}\rho_{2}\right),
\end{equation}
and
\begin{equation}
C_{3}=\frac{1}{2X^{2}}\left(2XC_{2}+\frac{1}{\sqrt{2X}}\rho_{1}{}_{X}\right),\quad C_{4}=\frac{1}{2X^{2}}\left(2XC_{1}-\frac{1}{\sqrt{2X}}\rho_{1}{}_{X}+\frac{1}{2}\rho_{2}\right),
\end{equation}
where the coefficients of the r.h.s. should be understood as functions
of $\left(\phi,\frac{1}{\sqrt{2X}}\right)$.

\section{The coefficients of gradient matrix \label{app_ceof}}

For completeness, the coefficients entering the gradient matrix in
eq. (\ref{III-B_G_zz})-(\ref{III-B_G_zd}) are listed below.

For the $G_{\zeta\zeta}$ terms, the coefficients are: 

\begin{equation}
G_{\zeta\zeta}^{\left(0\right)}\equiv-\left[18H^{2}M(\alpha_{\text{L}}+1)\bar{\rho}_{\text{m}}(w_{\text{m}}+1)\right]^{2}(\alpha_{\text{T}}+1),
\end{equation}
\begin{eqnarray}
G_{\zeta\zeta}^{\left(2\right)} & \equiv & 108H^{2}M^{2}(\alpha_{\text{L}}+1)\bar{\rho}_{\text{m}}(w_{\text{m}}+1)\Big\{-2M^{2}(\alpha_{\text{H}}+1)(\alpha_{\text{L}}+1)\dot{H}\nonumber \\
 &  & +2HM^{2}(\alpha_{\text{L}}+1)\dot{\alpha}_{\text{H}}+4H^{2}M^{2}(\alpha_{\text{L}}+1)(\alpha_{\text{H}}\alpha_{\text{M}}+\alpha_{\text{H}}+\alpha_{\text{M}}-\alpha_{\text{T}})\nonumber \\
 &  & +6H^{2}M^{2}(\alpha_{\text{H}}+1)(\alpha_{\text{L}}+1)c_{\text{m}}^{2}-\bar{\rho}_{\text{m}}(w_{\text{m}}+1)\left[(\alpha_{\text{H}}+1)^{2}-\alpha_{\text{A}}(\alpha_{\text{T}}+1)\right]\Big\},
\end{eqnarray}
\begin{eqnarray}
G_{\zeta\zeta}^{\left(4\right)} & \equiv & 9M^{2}\Big\{4H^{2}M^{2}(\alpha_{\text{L}}+1)\Big[-4M^{2}(\alpha_{\text{H}}+1)(\alpha_{\text{L}}+1)\dot{H}\bar{\rho}_{\text{m}}(w_{\text{m}}+1)\Big(2(\alpha_{\text{H}}+1)^{2}(\alpha_{\text{L}}-1)\nonumber \\
 &  & -\alpha_{\text{A}}\left(2(\alpha_{\text{H}}(\alpha_{\text{M}}+2)+(\alpha_{\text{L}}-1)\alpha_{\text{T}}+\alpha_{\text{L}}+\alpha_{\text{M}}+1)+3(\alpha_{\text{H}}+1)c_{\text{m}}^{2}\right)\Big)\Big]\nonumber \\
 &  & +\alpha_{\text{A}}\bar{\rho}_{\text{m}}(w_{\text{m}}+1)\left[\bar{\rho}_{\text{m}}(w_{\text{m}}+1)\left(2(\alpha_{\text{H}}+1)^{2}-\alpha_{\text{A}}(\alpha_{\text{T}}+1)\right)-4M^{2}(\alpha_{\text{H}}+1)(\alpha_{\text{L}}+1)\dot{H}\right]\nonumber \\
 &  & +16H^{3}M^{4}(\alpha_{\text{L}}+1)^{2}\dot{\alpha}_{\text{H}}+16H^{4}M^{4}(\alpha_{\text{L}}+1)^{2}(\alpha_{\text{H}}(\alpha_{\text{M}}+1)+\alpha_{\text{M}}-\alpha_{\text{T}})\nonumber \\
 &  & -4HM^{2}\bar{\rho}_{\text{m}}(w_{\text{m}}+1)\left(\alpha_{\text{A}}\left((\alpha_{\text{L}}+1)\dot{\alpha}_{\text{H}}+(\alpha_{\text{H}}+1)\dot{\alpha}_{\text{L}}\right)-(\alpha_{\text{H}}+1)(\alpha_{\text{L}}+1)\dot{\alpha}_{\text{A}}\right)\Big\},
\end{eqnarray}
\begin{eqnarray}
G_{\zeta\zeta}^{\left(6\right)} & \equiv & 12M^{4}\Big\{\alpha_{\text{A}}\alpha_{\text{L}}\left(2M^{2}(\alpha_{\text{H}}+1)(\alpha_{\text{L}}+1)\dot{H}+\bar{\rho}_{\text{m}}(w_{\text{m}}+1)\left(\alpha_{\text{A}}(\alpha_{\text{T}}+1)-2(\alpha_{\text{H}}+1)^{2}\right)\right)\nonumber \\
 &  & +2HM^{2}\left(\alpha_{\text{A}}\left(\alpha_{\text{L}}(\alpha_{\text{L}}+1)\dot{\alpha}_{\text{H}}-(\alpha_{\text{H}}+1)\dot{\alpha}_{\text{L}}\right)-(\alpha_{\text{H}}+1)\alpha_{\text{L}}(\alpha_{\text{L}}+1)\dot{\alpha}_{\text{A}}\right)\nonumber \\
 &  & +2H^{2}M^{2}\alpha_{\text{L}}(\alpha_{\text{L}}+1)\left(\alpha_{\text{A}}(\alpha_{\text{H}}(\alpha_{\text{M}}+3)+\alpha_{\text{M}}-2\alpha_{\text{T}}+1)+2(\alpha_{\text{H}}+1)^{2}\right)\Big\},
\end{eqnarray}
and
\begin{equation}
G_{\zeta\zeta}^{\left(8\right)}\equiv4M^{6}\alpha_{\text{A}}\alpha_{\text{L}}^{2}\left[2(\alpha_{\text{H}}+1)^{2}-\alpha_{\text{A}}(\alpha_{\text{T}}+1)\right].
\end{equation}

For the $G_{\delta\delta}$ terms, the coefficients are: 

\begin{eqnarray}
G_{\delta\delta}^{\left(0\right)} & \equiv & -54a^{2}H^{2}M^{4}(\alpha_{\text{L}}+1)\bar{\rho}_{\text{m}}^{2}(w_{\text{m}}+1)\Big\{12H^{2}(\alpha_{\text{L}}+1)c_{\text{m}}^{2}\left(H^{2}(-\alpha_{\text{M}}+3w_{\text{m}}-6)-\dot{H}\right)\nonumber \\
 &  & +M^{-4}\bar{\rho}_{\text{m}}(w_{\text{m}}+1)\left(2M^{2}(\alpha_{\text{L}}+1)\dot{H}+\bar{\rho}_{\text{m}}(w_{\text{m}}+1)\right)+12H^{2}(\alpha_{\text{L}}+1)w_{\text{m}}\dot{H}\nonumber \\
 &  & -24H^{3}(\alpha_{\text{L}}+1)c_{\text{m}}\dot{c}_{\text{m}}-36H^{4}(\alpha_{\text{L}}+1)c_{\text{m}}^{4}+12H^{4}(\alpha_{\text{L}}+1)(\alpha_{\text{M}}+6)w_{\text{m}}\Big\},
\end{eqnarray}
\begin{eqnarray}
G_{\delta\delta}^{\left(2\right)} & \equiv & 9a^{2}\bar{\rho}_{\text{m}}\Big\{6H^{2}M^{2}(\alpha_{\text{L}}+1)c_{\text{m}}^{2}\Big[2H^{2}M^{2}\Big(4M^{2}(\alpha_{\text{L}}+1)\dot{H}\nonumber \\
 &  & -\alpha_{\text{A}}\bar{\rho}_{\text{m}}(w_{\text{m}}+1)(-\alpha_{\text{L}}(\alpha_{\text{M}}+4)+\alpha_{\text{M}}+3(\alpha_{\text{L}}-1)w_{\text{m}}+8)\Big)\nonumber \\
 &  & -\alpha_{\text{A}}\bar{\rho}_{\text{m}}(w_{\text{m}}+1)\left(2M^{2}(\alpha_{\text{L}}+3)\dot{H}+\bar{\rho}_{\text{m}}(w_{\text{m}}+1)\right)\nonumber \\
 &  & +40H^{4}M^{4}(\alpha_{\text{L}}+1)+2HM^{2}(\alpha_{\text{L}}+1)\bar{\rho}_{\text{m}}(w_{\text{m}}+1)\dot{\alpha}_{\text{A}}\Big]\nonumber \\
 &  & -4H^{4}M^{4}(\alpha_{\text{L}}+1)\Big(12M^{2}(\alpha_{\text{L}}+1)w_{\text{m}}\dot{H}+\bar{\rho}_{\text{m}}(w_{\text{m}}+1)(2(\alpha_{\text{L}}+1)(\alpha_{\text{M}}+1)\nonumber \\
 &  & +3\alpha_{\text{A}}w_{\text{m}}(\alpha_{\text{L}}(\alpha_{\text{M}}+4)-\alpha_{\text{M}}-8))\Big)+4H^{2}M^{2}(\alpha_{\text{L}}+1)\bar{\rho}_{\text{m}}(w_{\text{m}}+1)\nonumber \\
 &  & \times\Big(M^{2}\dot{H}(3(\alpha_{\text{L}}+3)\alpha_{\text{A}}w_{\text{m}}-2(\alpha_{\text{L}}+1))+\bar{\rho}_{\text{m}}(w_{\text{m}}+1)(\alpha_{\text{L}}-\alpha_{\text{A}}-1)\Big)\nonumber \\
 &  & +\alpha_{\text{A}}\bar{\rho}_{\text{m}}^{2}(w_{\text{m}}+1)^{2}\left(\bar{\rho}_{\text{m}}(w_{\text{m}}+1)-2M^{2}(\alpha_{\text{L}}+1)\dot{H}\right)\nonumber \\
 &  & -2HM^{2}\bar{\rho}_{\text{m}}^{2}(w_{\text{m}}+1)^{2}\left(\alpha_{\text{A}}\dot{\alpha}_{\text{L}}-(\alpha_{\text{L}}+1)\dot{\alpha}_{\text{A}}\right)-240H^{6}M^{6}(\alpha_{\text{L}}+1)^{2}w_{\text{m}}\nonumber \\
 &  & +24H^{3}M^{4}(\alpha_{\text{L}}+1)c_{\text{m}}\dot{c}_{\text{m}}\left(4H^{2}M^{2}(\alpha_{\text{L}}+1)+(\alpha_{\text{L}}-1)\alpha_{\text{A}}\bar{\rho}_{\text{m}}(w_{\text{m}}+1)\right)\nonumber \\
 &  & +12H^{3}M^{4}\bar{\rho}_{\text{m}}(w_{\text{m}}+1)(\alpha_{\text{L}}+1)\left[3H\left(\alpha_{\text{L}}-1\right)\alpha_{\text{A}}c_{\text{m}}^{4}-(\alpha_{\text{L}}+1)w_{\text{m}}\dot{\alpha}_{\text{A}}\right]\Big\},
\end{eqnarray}
\begin{eqnarray}
G_{\delta\delta}^{\left(4\right)} & \equiv & -3a^{2}M^{2}\bar{\rho}_{\text{m}}\Big\{-3c_{\text{m}}^{2}\Big[2H^{2}M^{2}\alpha_{\text{A}}\Big(8M^{2}\alpha_{\text{L}}(\alpha_{\text{L}}+1)\dot{H}\nonumber \\
 &  & +\bar{\rho}_{\text{m}}(w_{\text{m}}+1)(\alpha_{\text{L}}(\alpha_{\text{A}}(-\alpha_{\text{M}}+3w_{\text{m}}-6)-4)-4)\Big)\nonumber \\
 &  & +\alpha_{\text{A}}^{2}\bar{\rho}_{\text{m}}(w_{\text{m}}+1)\left(\bar{\rho}_{\text{m}}(w_{\text{m}}+1)-2M^{2}\alpha_{\text{L}}\dot{H}\right)\nonumber \\
 &  & +16H^{4}M^{4}(\alpha_{\text{L}}+1)(5\alpha_{\text{L}}\alpha_{\text{A}}+\alpha_{\text{L}}+1)-2HM^{2}\alpha_{\text{A}}^{2}\bar{\rho}_{\text{m}}(w_{\text{m}}+1)\dot{\alpha}_{\text{L}}\Big]\nonumber \\
 &  & +2\Big[120H^{4}M^{4}\alpha_{\text{L}}(\alpha_{\text{L}}+1)\alpha_{\text{A}}w_{\text{m}}+H^{2}M^{2}\alpha_{\text{L}}\Big(24M^{2}(\alpha_{\text{L}}+1)\alpha_{\text{A}}w_{\text{m}}\dot{H}\nonumber \\
 &  & -\bar{\rho}_{\text{m}}(w_{\text{m}}+1)\left(-2(\alpha_{\text{L}}+1)(\alpha_{\text{M}}-1)\alpha_{\text{A}}+4(\alpha_{\text{L}}+1)+3(\alpha_{\text{M}}+6)\alpha_{\text{A}}^{2}w_{\text{m}}\right)\Big)\nonumber \\
 &  & +\alpha_{\text{L}}\alpha_{\text{A}}\bar{\rho}_{\text{m}}(w_{\text{m}}+1)\left(2\bar{\rho}_{\text{m}}(w_{\text{m}}+1)-M^{2}\dot{H}(2\alpha_{\text{L}}+3\alpha_{\text{A}}w_{\text{m}}+2)\right)\nonumber \\
 &  & +HM^{2}\bar{\rho}_{\text{m}}(w_{\text{m}}+1)\left(2\alpha_{\text{L}}(\alpha_{\text{L}}+1)\dot{\alpha}_{\text{A}}+\alpha_{\text{A}}\dot{\alpha}_{\text{L}}(2-3\alpha_{\text{A}}w_{\text{m}})\right)\Big]\nonumber \\
 &  & +12HM^{2}\alpha_{\text{L}}\alpha_{\text{A}}c_{\text{m}}\dot{c}_{\text{m}}\left(\alpha_{\text{A}}\bar{\rho}_{\text{m}}(w_{\text{m}}+1)-8H^{2}M^{2}(\alpha_{\text{L}}+1)\right)\nonumber \\
 &  & +18H^{2}M^{2}\alpha_{\text{L}}\alpha_{\text{A}}^{2}c_{\text{m}}^{4}\bar{\rho}_{\text{m}}(w_{\text{m}}+1)\Big\},
\end{eqnarray}
\begin{eqnarray}
G_{\delta\delta}^{\left(6\right)} & \equiv & 4a^{2}M^{4}\alpha_{\text{L}}\alpha_{\text{A}}\bar{\rho}_{\text{m}}\Big\{3c_{\text{m}}^{2}\left[\alpha_{\text{A}}\left(M^{2}\alpha_{\text{L}}\dot{H}-\bar{\rho}_{\text{m}}(w_{\text{m}}+1)\right)+H^{2}M^{2}(\alpha_{\text{L}}(5\alpha_{\text{A}}+4)+4)\right]\nonumber \\
 &  & +\alpha_{\text{L}}\left[w_{\text{m}}\left(\bar{\rho}_{\text{m}}-3M^{2}\alpha_{\text{A}}\left(\dot{H}+5H^{2}\right)\right)+\bar{\rho}_{\text{m}}\right]+6HM^{2}\alpha_{\text{L}}\alpha_{\text{A}}c_{\text{m}}\dot{c}_{\text{m}}\Big\},
\end{eqnarray}
and
\begin{equation}
G_{\delta\delta}^{\left(8\right)}\equiv\left(2aM^{3}\alpha_{\text{L}}\alpha_{\text{A}}c_{\text{m}}\right)^{2}\bar{\rho}_{\text{m}}.
\end{equation}

For the $G_{\zeta\delta}$ terms, the coefficients are: 

\begin{eqnarray}
G_{\zeta\delta}^{\left(0\right)} & \equiv & 972aH^{4}M^{2}(\alpha_{\text{L}}+1)^{2}\bar{\rho}_{\text{m}}^{2}(w_{\text{m}}+1)\Big\{ c_{\text{m}}^{2}\left[\dot{H}+H^{2}(\alpha_{\text{M}}-3w_{\text{m}}+6)\right]\nonumber \\
 &  & -w_{\text{m}}\left(\dot{H}+H^{2}(\alpha_{\text{M}}+6)\right)+2Hc_{\text{m}}\dot{c}_{\text{m}}+3H^{2}c_{\text{m}}^{4}\Big\},
\end{eqnarray}
\begin{eqnarray}
G_{\zeta\delta}^{\left(2\right)} & \equiv & -108aH^{2}(\alpha_{\text{L}}+1)\bar{\rho}_{\text{m}}\Big\{-3M^{2}c_{\text{m}}^{2}\Big[H^{2}\Big(2M^{2}(\alpha_{\text{L}}+1)\dot{H}+\bar{\rho}_{\text{m}}(w_{\text{m}}+1)(\alpha_{\text{H}}(\alpha_{\text{L}}+1)\nonumber \\
 &  & +\alpha_{\text{A}}(-\alpha_{\text{M}}+3w_{\text{m}}-6))\Big)-\alpha_{\text{A}}\bar{\rho}_{\text{m}}(w_{\text{m}}+1)\dot{H}+10H^{4}M^{2}(\alpha_{\text{L}}+1)\Big]\nonumber \\
 &  & +H^{2}M^{2}\big[+6M^{2}(\alpha_{\text{L}}+1)w_{\text{m}}\dot{H}+\bar{\rho}_{\text{m}}(w_{\text{m}}+1)((\alpha_{\text{L}}+1)\nonumber \\
 &  & \times((\alpha_{\text{H}}+2)\alpha_{\text{M}}+\alpha_{\text{H}}+3\alpha_{\text{H}}w_{\text{m}}+2)-3(\alpha_{\text{M}}+6)\alpha_{\text{A}}w_{\text{m}})\Big]\nonumber \\
 &  & -\bar{\rho}_{\text{m}}(w_{\text{m}}+1)\left(M^{2}\dot{H}((\alpha_{\text{H}}+2)(\alpha_{\text{L}}+1)+3\alpha_{\text{A}}w_{\text{m}})+(\alpha_{\text{H}}+1)\bar{\rho}_{\text{m}}(w_{\text{m}}+1)\right)\nonumber \\
 &  & -6HM^{2}c_{\text{m}}\dot{c}_{\text{m}}\left(2H^{2}M^{2}(\alpha_{\text{L}}+1)-\alpha_{\text{A}}\bar{\rho}_{\text{m}}(w_{\text{m}}+1)\right)+30H^{4}M^{4}(\alpha_{\text{L}}+1)w_{\text{m}}\nonumber \\
 &  & +9H^{2}M^{2}\alpha_{\text{A}}c_{\text{m}}^{4}\bar{\rho}_{\text{m}}(w_{\text{m}}+1)+HM^{2}(\alpha_{\text{L}}+1)\bar{\rho}_{\text{m}}(w_{\text{m}}+1)\dot{\alpha}_{\text{H}}\Big\},
\end{eqnarray}
\begin{eqnarray}
G_{\zeta\delta}^{\left(4\right)} & \equiv & 9a\bar{\rho}_{\text{m}}\Big\{3M^{2}c_{\text{m}}^{2}\Big[H^{2}\alpha_{\text{A}}\Big(4M^{2}(\alpha_{\text{L}}+1)(3\alpha_{\text{L}}+1)\dot{H}-\bar{\rho}_{\text{m}}(w_{\text{m}}+1)(2\alpha_{\text{H}}(\alpha_{\text{L}}+1)\nonumber \\
 &  & +\alpha_{\text{A}}(-\alpha_{\text{M}}+3w_{\text{m}}-6))\Big)+\alpha_{\text{A}}^{2}\bar{\rho}_{\text{m}}(w_{\text{m}}+1)\dot{H}+4H^{4}M^{2}(\alpha_{\text{L}}+1)(4(\alpha_{\text{H}}+1)(\alpha_{\text{L}}+1)\nonumber \\
 &  & +(7\alpha_{\text{L}}-3)\alpha_{\text{A}})-4H^{3}M^{2}(\alpha_{\text{L}}+1)^{2}\dot{\alpha}_{\text{A}}\Big]+H^{2}M^{2}\Big[4M^{2}(\alpha_{\text{L}}+1)\dot{H}(2(\alpha_{\text{H}}+2)(\alpha_{\text{L}}+1)\nonumber \\
 &  & -3(3\alpha_{\text{L}}+1)\alpha_{\text{A}}w_{\text{m}})+\bar{\rho}_{\text{m}}(w_{\text{m}}+1)\Big(-8(\alpha_{\text{H}}+1)\left(\alpha_{\text{L}}^{2}-1\right)\nonumber \\
 &  & +2(\alpha_{\text{L}}+1)\alpha_{\text{A}}((\alpha_{\text{H}}+2)\alpha_{\text{M}}+3\alpha_{\text{H}}(w_{\text{m}}+1)+6)-3(\alpha_{\text{M}}+6)\alpha_{\text{A}}^{2}w_{\text{m}}\Big)\Big]\nonumber \\
 &  & +\alpha_{\text{A}}\bar{\rho}_{\text{m}}(w_{\text{m}}+1)\left(M^{2}\dot{H}(2(\alpha_{\text{H}}+2)(\alpha_{\text{L}}+1)-3\alpha_{\text{A}}w_{\text{m}})-2(\alpha_{\text{H}}+1)\bar{\rho}_{\text{m}}(w_{\text{m}}+1)\right)\nonumber \\
 &  & +6HM^{2}\alpha_{\text{A}}c_{\text{m}}\dot{c}_{\text{m}}\left(4H^{2}M^{2}\left(\alpha_{\text{L}}^{2}-1\right)+\alpha_{\text{A}}\bar{\rho}_{\text{m}}(w_{\text{m}}+1)\right)+9H^{2}M^{2}\alpha_{\text{A}}^{2}c_{\text{m}}^{4}\bar{\rho}_{\text{m}}(w_{\text{m}}+1)\nonumber \\
 &  & +2HM^{2}\bar{\rho}_{\text{m}}(w_{\text{m}}+1)\left(\alpha_{\text{A}}\left((\alpha_{\text{L}}+1)\dot{\alpha}_{\text{H}}+(\alpha_{\text{H}}+2)\dot{\alpha}_{\text{L}}\right)-(\alpha_{\text{H}}+2)(\alpha_{\text{L}}+1)\dot{\alpha}_{\text{A}}\right)-4H^{3}M^{4}\nonumber \\
 &  & \times(\alpha_{\text{L}}+1)\left[(\alpha_{\text{L}}+1)\left(2\dot{\alpha}_{\text{H}}-3w_{\text{m}}\dot{\alpha}_{\text{A}}\right)+3Hw_{\text{m}}(2\alpha_{\text{H}}(\alpha_{\text{L}}+1)+(7\alpha_{\text{L}}-3)\alpha_{\text{A}}\right])\Big\},
\end{eqnarray}
\begin{eqnarray}
G_{\zeta\delta}^{\left(6\right)} & \equiv & 6M^{2}a\bar{\rho}_{\text{m}}\Big\{\alpha_{\text{L}}\alpha_{\text{A}}\Big[-M^{2}\dot{H}\left(2(\alpha_{\text{H}}+2)(\alpha_{\text{L}}+1)+3\alpha_{\text{A}}\left(c_{\text{m}}^{2}-w_{\text{m}}\right)\right)\nonumber \\
 &  & +4(\alpha_{\text{H}}+1)\bar{\rho}_{\text{m}}(w_{\text{m}}+1)\Big]+HM^{2}\Big[2(\alpha_{\text{H}}+2)\alpha_{\text{L}}(\alpha_{\text{L}}+1)\dot{\alpha}_{\text{A}}\nonumber \\
 &  & +\alpha_{\text{A}}\left(\dot{\alpha}_{\text{L}}\left(2\alpha_{\text{H}}+3\alpha_{\text{A}}\left(c_{\text{m}}^{2}-w_{\text{m}}\right)+4\right)-2\alpha_{\text{L}}\left((\alpha_{\text{L}}+1)\dot{\alpha}_{\text{H}}+3\alpha_{\text{A}}c_{\text{m}}\dot{c}_{\text{m}}\right)\right)\Big]\nonumber \\
 &  & +H^{2}M^{2}\alpha_{\text{L}}\Big[-8(\alpha_{\text{H}}+1)(\alpha_{\text{L}}+1)+15\alpha_{\text{A}}^{2}\left(w_{\text{m}}-c_{\text{m}}^{2}\right)\nonumber \\
 &  & +2(\alpha_{\text{L}}+1)\alpha_{\text{A}}\left(6(\alpha_{\text{H}}+1)c_{\text{m}}^{2}-\alpha_{\text{H}}(3w_{\text{m}}+2)-4\right)\Big]\Big\},
\end{eqnarray}
and
\begin{equation}
G_{\zeta\delta}^{\left(8\right)}\equiv-8aM^{4}\alpha_{\text{L}}^{2}\alpha_{\text{A}}(\alpha_{\text{H}}+1)\bar{\rho}_{\text{m}}.
\end{equation}

\section{The series coefficients of high-$k$ expansions \label{App_series_coeff}}

The coefficients appearing in the high-$k$ expansions used in eq.
(\ref{III-D_I_zd}) are collected here.

\begin{eqnarray}
\mathcal{K}_{\zeta\delta}^{\left(-1\right)} & \equiv & -\frac{3a^{2}\bar{\rho}_{\text{m}}}{2\alpha_{\text{L}}},\quad\mathcal{G}_{\zeta\delta}^{\left(-1\right)}\equiv-\frac{a^{2}\left(1+\alpha_{\text{H}}\right)\bar{\rho}_{\text{m}}}{\alpha_{\text{A}}},\\
\mathcal{I}_{\zeta\delta}^{\left(-2\right)} & \equiv & \frac{3a^{3}H\bar{\rho}_{\text{m}}}{2\alpha_{\text{L}}\alpha_{\text{A}}}\left[2\alpha_{\text{H}}(\alpha_{\text{L}}+1)+3\alpha_{\text{A}}\left(w_{\text{m}}-c_{\text{m}}^{2}\right)\right],\\
\mathcal{G}_{\delta\delta}^{\left(-2\right)} & \equiv & a^{4}\bar{\rho}_{\text{m}}\left[\frac{\bar{\rho}_{\text{m}}}{2M^{2}\alpha_{\text{A}}}+\frac{3c_{\text{m}}\dot{c}_{\text{m}}}{1+w_{\text{m}}}H+\frac{3}{2}\frac{c_{\text{m}}^{2}-w_{\text{m}}}{1+w_{\text{m}}}\left(5H^{2}+\dot{H}\right)\right],
\end{eqnarray}
and
\begin{eqnarray}
\mathcal{G}_{\zeta\zeta}^{\left(-2\right)} & \equiv & \frac{6a^{2}M^{2}}{\alpha_{\text{L}}^{2}\alpha_{\text{A}}}\Big\{\left(1+\alpha_{\text{H}}\right)\left(1+\alpha_{\text{L}}\right)\left[\left(\left(3+\alpha_{\text{M}}\right)\alpha_{\text{A}}-2\left(1+\alpha_{\text{H}}\right)\right)H^{2}+\dot{H}\right]\nonumber \\
 &  & +\left[\alpha_{\text{L}}\left(1+\alpha_{\text{L}}\right)\left(\alpha_{\text{A}}\dot{\alpha}_{\text{H}}-\left(1+\alpha_{\text{H}}\right)\dot{\alpha}_{\text{A}}\right)-\left(1+\alpha_{\text{H}}\right)\alpha_{\text{A}}\dot{\alpha}_{\text{L}}\right]H\Big\}.
\end{eqnarray}

\section{The coefficients of constraint equations\label{App_constr_eq}}

The coefficients of the constraint equations (\ref{III-F_eom_A})
and (\ref{III-F_eom_B}) are given here.

\begin{eqnarray}
\mathcal{E}_{A\dot{\delta}} & \equiv & -\Xi^{-1}6H(\alpha_{\text{L}}+1)\bar{\rho}_{\text{m}},\quad\mathcal{E}_{A\dot{\zeta}}\equiv\Xi^{-1}12HM^{2}(\alpha_{\text{L}}+1)\frac{\mathit{k}^{2}}{a^{2}},\\
\mathcal{E}_{A\zeta} & \equiv & \Xi^{-1}2\frac{\mathit{k}^{2}}{a^{2}}(\alpha_{\text{H}}+1)\left[3\bar{\rho}_{\text{m}}(w_{\text{m}}+1)-2\frac{\mathit{k}^{2}}{a^{2}}M^{2}\alpha_{\text{L}}\right],\\
\mathcal{E}_{A\delta} & \equiv & \Xi^{-1}\bar{\rho}_{\text{m}}\left\{ 2M^{2}\left[\frac{\mathit{k}^{2}}{a^{2}}\alpha_{L}-9H^{2}(\alpha_{L}+1)\left(c_{\text{m}}^{2}-w_{\text{m}}\right)\right]-3\bar{\rho}_{\text{m}}(w_{\text{m}}+1)\right\} ,
\end{eqnarray}
and

\begin{eqnarray}
\mathcal{E}_{B\dot{\delta}} & \equiv & \left(a\Xi\right)^{-1}3\bar{\rho}_{\text{m}}\left(\alpha_{A}-6\frac{a^{2}}{k^{2}}H^{2}(\alpha_{\text{L}}+1)\right),\\
\mathcal{E}_{B\zeta} & \equiv & -\left(a\Xi\right)^{-1}12\frac{\mathit{k}^{2}}{a^{2}}HM^{2}(\alpha_{\text{H}}+1)(\alpha_{\text{L}}+1),\\
\mathcal{E}_{B\delta} & \equiv & \left(a\Xi\right)^{-1}3H\bar{\rho}_{\text{m}}\left\{ 3\left(c_{\text{m}}^{2}-w_{\text{m}}\right)\left[\alpha_{\text{A}}-6\frac{a^{2}}{k^{2}}H^{2}(\alpha_{\text{L}}+1)\right]+2(\alpha_{\text{L}}+1)\right\} ,\\
\mathcal{E}_{B\dot{\zeta}} & \equiv & \left(a\Xi\right)^{-1}\left\{ \frac{\mathit{k}^{2}}{a^{2}}6M^{2}\alpha_{\text{A}}(\alpha_{\text{L}}+1)-9\bar{\rho}_{\text{m}}(w_{\text{m}}+1)\left[\frac{a^{2}}{k^{2}}6H^{2}(\alpha_{\text{L}}+1)-\alpha_{\text{A}}\right]\right\} ,
\end{eqnarray}
where $\Xi$ is defined in eq. (\ref{III-B_Xi}).

\acknowledgments
Z. Yao thanks A. Silvestri for useful discussions. This work is supported by the European Research Council under the H2020 ERC Consolidator Grant “Gravitational Physics from the Universe Large scales Evolution” (Grant No. 101126217 — GraviPULSE).


\begin{thebibliography}{99}

\bibitem{Slosar:2019flp}
A.~Slosar et~al., \emph{{Dark Energy and Modified Gravity}},
  \href{https://arxiv.org/abs/1903.12016}{{\ttfamily 1903.12016}}.

\bibitem{Collaboration:2025ac}
D.~Collaboration, A.~G. Adame, J.~Aguilar, S.~Ahlen, S.~Alam, D.~M. Alexander
  et~al., \emph{DESI 2024 VI: Cosmological constraints from the measurements of
  baryon acoustic oscillations},
  \href{https://doi.org/10.1088/1475-7516/2025/02/021}{\emph{JCAP}
  {\bfseries 02} (2025) 021}
  [\href{https://arxiv.org/abs/2404.03002}{{\ttfamily 2404.03002}}].

\bibitem{Calderon:2024aa}
R.~Calderon, K.~Lodha, A.~Shafieloo, E.~Linder, W.~Sohn, A.~de~Mattia et~al.,
  \emph{DESI 2024: Reconstructing dark energy using crossing statistics with
  DESI DR1 BAO data},
  \href{https://doi.org/10.1088/1475-7516/2024/10/048}{\emph{JCAP} {\bfseries 10} (2024) 048} [\href{https://arxiv.org/abs/2405.04216}{{\ttfamily 2405.04216}}].

\bibitem{Lodha:2025aa}
K.~Lodha, R.~Calderon, W.~L. Matthewson, A.~Shafieloo, M.~Ishak, J.~Pan et~al.,
  \emph{Extended dark energy analysis using DESI DR2 BAO measurements},
  \href{https://arxiv.org/abs/2503.14743}{{\ttfamily 2503.14743}}.

\bibitem{Louis:2025aa}
T.~Louis, A.~L. Posta, Z.~Atkins, H.~T. Jense, I.~Abril-Cabezas, G.~E. Addison
  et~al., \emph{The atacama cosmology telescope: Dr6 power spectra, likelihoods
  and $\Lambda$CDM parameters},
  \href{https://arxiv.org/abs/2503.14452}{{\ttfamily 2503.14452}}.

\bibitem{AtacamaCosmologyTelescope:2025nti}
{\scshape Atacama Cosmology Telescope} collaboration, E.~Calabrese et~al.,
  \emph{{The Atacama Cosmology Telescope: DR6 constraints on extended
  cosmological models}},
  \href{https://doi.org/10.1088/1475-7516/2025/11/063}{\emph{JCAP} {\bfseries
  11} (2025) 063} [\href{https://arxiv.org/abs/2503.14454}{{\ttfamily
  2503.14454}}].

\bibitem{Collaboration:2024aa}
D.~Collaboration, T.~M.~C. Abbott, M.~Acevedo, M.~Aguena, A.~Alarcon, S.~Allam
  et~al., \emph{The dark energy survey: Cosmology results with ~1500 new
  high-redshift type ia supernovae using the full 5-year dataset},
  {\emph{FERMILAB-PUB-23-0821-PPD} (2024) }
  [\href{https://arxiv.org/abs/2401.02929}{{\ttfamily 2401.02929}}].

\bibitem{Ormondroyd:2025aa}
A.~Ormondroyd, W.~Handley, M.~Hobson and A.~Lasenby, \emph{Comparison of
  dynamical dark energy with $\Lambda$CDM in light of DESI DR2},
  \href{https://arxiv.org/abs/2503.17342}{{\ttfamily 2503.17342}}.

\bibitem{Gu:2025aa}
G.~Gu, X.~Wang, Y.~Wang, G.-B. Zhao, L.~Pogosian, K.~Koyama et~al.,
  \emph{Dynamical dark energy in light of the DESI DR2 baryonic acoustic
  oscillations measurements},
  \href{https://arxiv.org/abs/2504.06118}{{\ttfamily 2504.06118}}.

\bibitem{Akarsu:2024eoo}
{\"O}.~Akarsu, A.~De~Felice, E.~Di~Valentino, S.~Kumar, R.~C. Nunes,
  E.~{\"O}z{\"u}lker et~al., \emph{{Cosmological constraints on
  {\ensuremath{\Lambda}}sCDM scenario in a type II minimally modified
  gravity}}, \href{https://doi.org/10.1103/PhysRevD.110.103527}{\emph{Phys.
  Rev. D} {\bfseries 110} (2024) 103527}
  [\href{https://arxiv.org/abs/2406.07526}{{\ttfamily 2406.07526}}].

\bibitem{Orchard:2024bve}
L.~Orchard and V.~H. C{\'a}rdenas, \emph{{Probing dark energy evolution
  post-DESI 2024}},
  \href{https://doi.org/10.1016/j.dark.2024.101678}{\emph{Phys. Dark Univ.}
  {\bfseries 46} (2024) 101678}
  [\href{https://arxiv.org/abs/2407.05579}{{\ttfamily 2407.05579}}].

\bibitem{Chudaykin:2024gol}
A.~Chudaykin and M.~Kunz, \emph{{Modified gravity interpretation of the
  evolving dark energy in light of DESI data}},
  \href{https://doi.org/10.1103/PhysRevD.110.123524}{\emph{Phys. Rev. D}
  {\bfseries 110} (2024) 123524}
  [\href{https://arxiv.org/abs/2407.02558}{{\ttfamily 2407.02558}}].

\bibitem{Alestas:2024gxe}
G.~Alestas, M.~Delgado, I.~Ruiz, Y.~Akrami, M.~Montero and S.~Nesseris,
  \emph{{Is curvature-assisted quintessence observationally viable?}},
  \href{https://doi.org/10.1103/PhysRevD.110.106010}{\emph{Phys. Rev. D}
  {\bfseries 110} (2024) 106010}
  [\href{https://arxiv.org/abs/2406.09212}{{\ttfamily 2406.09212}}].

\bibitem{Wang:2024dka}
H.~Wang and Y.-S. Piao, \emph{{Dark energy in light of DESI DR1 and Hubble
  tension}}, \href{https://doi.org/10.1016/j.physletb.2026.140180}{\emph{Phys.
  Lett. B} {\bfseries 873} (2026) 140180}
  [\href{https://arxiv.org/abs/2404.18579}{{\ttfamily 2404.18579}}].

\bibitem{Hogas:2025ahb}
M.~H{\"o}g{\r{a}}s and E.~M{\"o}rtsell, \emph{{Bimetric gravity improves the
  fit to DESI BAO and eases the Hubble tension}},
  \href{https://doi.org/10.1103/zz5k-kzzk}{\emph{Phys. Rev. D} {\bfseries 112}
  (2025) 103515} [\href{https://arxiv.org/abs/2507.03743}{{\ttfamily
  2507.03743}}].

\bibitem{Gonzalez-Fuentes:2026rgu}
A.~Gonz{\'a}lez-Fuentes and A.~G{\'o}mez-Valent, \emph{{Exploring the interplay
  of late-time dynamical dark energy and new physics before recombination}},
  \href{https://doi.org/10.1088/1475-7516/2026/07/040}{\emph{JCAP} {\bfseries
  07} (2026) 040} [\href{https://arxiv.org/abs/2603.26560}{{\ttfamily
  2603.26560}}].

\bibitem{Calderon:2026hbr}
R.~Calderon and E.~V. Linder, \emph{{Charging Across the Phantom Divide with
  Modified Gravity}},  \href{https://arxiv.org/abs/2605.26259}{{\ttfamily
  2605.26259}}.

\bibitem{Chanda:2026dmx}
P.~Chanda, S.~Das and S.~Das, \emph{{Dissipative Dark Energy can explain the
  DESI phantom crossing}},  \href{https://arxiv.org/abs/2606.04886}{{\ttfamily
  2606.04886}}.

\bibitem{Garcia-Garcia:2026nzy}
C.~Garc{\'\i}a-Garc{\'\i}a, P.~G. Ferreira and W.~J. Wolf, \emph{{The Status of
  Single Scalar Field Dark Energy}},
  \href{https://arxiv.org/abs/2607.07777}{{\ttfamily 2607.07777}}.

\bibitem{Gomez-Valent:2026ept}
A.~G{\'o}mez-Valent, Z.~Zheng and L.~Amendola, \emph{{Constraints on Coupled
  Dark Energy in the DESI Era}},
  \href{https://arxiv.org/abs/2604.12032}{{\ttfamily 2604.12032}}.

\bibitem{Ladeira:2026jne}
A.~Ladeira, R.~C. Nunes, S.~Pan and W.~Yang, \emph{{Joint constraints on
  neutrinos and dynamical dark energy in minimally modified gravity}},
  \href{https://doi.org/10.1103/q1ng-2xvp}{\emph{Phys. Rev. D} {\bfseries 113}
  (2026) 083503} [\href{https://arxiv.org/abs/2601.02077}{{\ttfamily
  2601.02077}}].

\bibitem{Efstratiou:2025iqi}
D.~Efstratiou, E.~A. Paraskevas and L.~Perivolaropoulos, \emph{{Addressing the
  DESI DR2 phantom-crossing anomaly and enhanced H0 tension with reconstructed
  scalar-tensor gravity}}, \href{https://doi.org/10.1103/zdcg-4sdf}{\emph{Phys.
  Rev. D} {\bfseries 113} (2026) 123525}
  [\href{https://arxiv.org/abs/2511.04610}{{\ttfamily 2511.04610}}].

\bibitem{Liu:2025bss}
R.~Liu, Y.~Zhu, W.~Hu and V.~Miranda, \emph{{Phantom mirage from axion dark
  energy}}, \href{https://doi.org/10.1103/3s1m-9zpc}{\emph{Phys. Rev. D}
  {\bfseries 113} (2026) 083506}
  [\href{https://arxiv.org/abs/2510.14957}{{\ttfamily 2510.14957}}].

\bibitem{Wolf:2025acj}
W.~J. Wolf, P.~G. Ferreira and C.~Garc{\'\i}a-Garc{\'\i}a, \emph{{Cosmological
  constraints on Galileon dark energy with broken shift symmetry}},
  \href{https://doi.org/10.1103/bxvj-bsv1}{\emph{Phys. Rev. D} {\bfseries 113}
  (2026) 023551} [\href{https://arxiv.org/abs/2509.17586}{{\ttfamily
  2509.17586}}].

\bibitem{Alam:2025epg}
S.~Alam and M.~W. Hossain, \emph{{Beyond CPL: Evidence for dynamical dark
  energy in three-parameter models}},
  \href{https://doi.org/10.1088/1475-7516/2026/04/042}{\emph{JCAP} {\bfseries
  04} (2026) 042} [\href{https://arxiv.org/abs/2510.03779}{{\ttfamily
  2510.03779}}].

\bibitem{Pookkillath:2026wyg}
M.~C. Pookkillath and S.~Tsujikawa, \emph{{Phantom-divide crossing and
  suppressed structure growth in kinetically braided dark energy with momentum
  exchange}},  \href{https://arxiv.org/abs/2607.26447}{{\ttfamily 2607.26447}}.

\bibitem{Khoury:2026svx}
J.~Khoury, M.-X. Lin and M.~Trodden, \emph{{Cosmological Evidence for Dark
  Axion-Dark Baryon Interactions from Apparent Phantom Crossing}},
  \href{https://arxiv.org/abs/2607.16191}{{\ttfamily 2607.16191}}.

\bibitem{Caldwell:2002aa}
R.~Caldwell, \emph{A phantom menace? cosmological consequences of a dark energy
  component with super-negative equation of state},
  \href{https://doi.org/10.1016/S0370-2693%2802%2902589-3}{\emph{Phys. Lett. B}
  {\bfseries 545} (2002) 23}
  [\href{https://arxiv.org/abs/astro-ph/9908168}{{\ttfamily
  astro-ph/9908168}}].

\bibitem{Feng:2005aa}
B.~Feng, X.~Wang and X.~Zhang, \emph{Dark energy constraints from the cosmic
  age and supernova},
  \href{https://doi.org/10.1016/j.physletb.2004.12.071}{\emph{Phys. Lett. B}
  {\bfseries 607} (2005) 35}
  [\href{https://arxiv.org/abs/astro-ph/0404224}{{\ttfamily
  astro-ph/0404224}}].

\bibitem{Piazza2004}
F.~Piazza and S.~Tsujikawa, \emph{Dilatonic ghost condensate as dark energy},
  \href{https://doi.org/10.1088/1475-7516/2004/07/004}{\emph{JCAP} {\bfseries
  07} (2004) 004} [\href{https://arxiv.org/abs/hep-th/0405054}{{\ttfamily
  hep-th/0405054}}].

\bibitem{Gannouji:2006aa}
R.~Gannouji, D.~Polarski, A.~Ranquet and A.~A. Starobinsky, \emph{Scalar-tensor
  models of normal and phantom dark energy},
  \href{https://doi.org/10.1088/1475-7516/2006/09/016}{\emph{JCAP}
  {\bfseries 0609} (2006) 016}
  [\href{https://arxiv.org/abs/astro-ph/0606287}{{\ttfamily
  astro-ph/0606287}}].

\bibitem{Amendola:2008aa}
L.~Amendola and S.~Tsujikawa, \emph{Phantom crossing, equation-of-state
  singularities, and local gravity constraints in f(r) models},
  \href{https://doi.org/10.1016/j.physletb.2007.12.041}{\emph{Phys. Lett. B}
  {\bfseries 660} (2008) 125}
  [\href{https://arxiv.org/abs/0705.0396}{{\ttfamily 0705.0396}}].

\bibitem{Ye:2025aa}
G.~Ye, M.~Martinelli, B.~Hu and A.~Silvestri, \emph{Hints of nonminimally
  coupled gravity in DESI 2024 baryon acoustic oscillation measurements},
  \href{https://doi.org/10.1103/PhysRevLett.134.181002}{\emph{Phys. Rev. Lett.}
  {\bfseries 134} (2025) 181002}
  [\href{https://arxiv.org/abs/2407.15832}{{\ttfamily 2407.15832}}].

\bibitem{Creminelli:2008wc}
P.~Creminelli, G.~D'Amico, J.~Norena and F.~Vernizzi, \emph{{The Effective
  Theory of Quintessence: the w<-1 Side Unveiled}},
  \href{https://doi.org/10.1088/1475-7516/2009/02/018}{\emph{JCAP} {\bfseries
  0902} (2009) 018} [\href{https://arxiv.org/abs/0811.0827}{{\ttfamily
  0811.0827}}].

\bibitem{Cai:2016aa}
Y.-F. Cai, S.~Capozziello, M.~D. Laurentis and E.~N. Saridakis, \emph{f(T)
  teleparallel gravity and cosmology},
  \href{https://doi.org/10.1088/0034-4885/79/10/106901}{\emph{Rept. Prog. Phys.}
  {\bfseries 79} (2016) 106901}
  [\href{https://arxiv.org/abs/1511.07586}{{\ttfamily 1511.07586}}].

\bibitem{Vikman:2005aa}
A.~Vikman, \emph{Can dark energy evolve to the phantom?},
  \href{https://doi.org/10.1103/PhysRevD.71.023515}{\emph{Phys.Rev.}
  {\bfseries D71} (2005) 023515}
  [\href{https://arxiv.org/abs/astro-ph/0407107}{{\ttfamily
  astro-ph/0407107}}].

\bibitem{ArmendarizPicon1999}
C.~Armendariz-Picon, T.~Damour and V.~F. Mukhanov, \emph{k - inflation},
  \href{https://doi.org/10.1016/S0370-2693(99)00603-6}{\emph{Phys. Lett. B}
  {\bfseries 458} (1999) 209}
  [\href{https://arxiv.org/abs/hep-th/9904075}{{\ttfamily hep-th/9904075}}].

\bibitem{Chiba2000}
T.~Chiba, T.~Okabe and M.~Yamaguchi, \emph{{Kinetically driven quintessence}},
  \href{https://doi.org/10.1103/PhysRevD.62.023511}{\emph{Phys. Rev. D}
  {\bfseries 62} (2000) 023511}
  [\href{https://arxiv.org/abs/astro-ph/9912463}{{\ttfamily
  astro-ph/9912463}}].

\bibitem{Horndeski1974}
G.~W. Horndeski, \emph{{Second-order scalar-tensor field equations in a
  four-dimensional space}},
  \href{https://doi.org/10.1007/BF01807638}{\emph{Int. J. Theor. Phys.}
  {\bfseries 10} (1974) 363}.

\bibitem{Deffayet2011}
C.~Deffayet, X.~Gao, D.~A. Steer and G.~Zahariade, \emph{{From k-essence to
  generalised Galileons}},
  \href{https://doi.org/10.1103/PhysRevD.84.064039}{\emph{Phys. Rev. D}
  {\bfseries 84} (2011) 064039}
  [\href{https://arxiv.org/abs/1103.3260}{{\ttfamily 1103.3260}}].

\bibitem{Matsumoto:2017qil}
J.~Matsumoto, \emph{{Phantom crossing dark energy in
  Horndeski{\textquoteright}s theory}},
  \href{https://doi.org/10.1103/PhysRevD.97.123538}{\emph{Phys. Rev. D}
  {\bfseries 97} (2018) 123538}
  [\href{https://arxiv.org/abs/1712.10015}{{\ttfamily 1712.10015}}].

\bibitem{Linder:2025pqt}
E.~V. Linder, \emph{{Cosmology after Phantom Crossing by Horndeski Gravity}},
  \href{https://arxiv.org/abs/2512.03139}{{\ttfamily 2512.03139}}.

\bibitem{Lin:2017oow}
C.~Lin and S.~Mukohyama, \emph{{A Class of Minimally Modified Gravity
  Theories}}, \href{https://doi.org/10.1088/1475-7516/2017/10/033}{\emph{JCAP}
  {\bfseries 1710} (2017) 033}
  [\href{https://arxiv.org/abs/1708.03757}{{\ttfamily 1708.03757}}].

\bibitem{DeFelice2020b}
A.~De~Felice, A.~Doll and S.~Mukohyama, \emph{A theory of type-ii minimally
  modified gravity},
  \href{https://doi.org/10.1088/1475-7516/2020/09/034}{\emph{JCAP} {\bfseries
  09} (2020) 034} [\href{https://arxiv.org/abs/2004.12549}{{\ttfamily
  2004.12549}}].

\bibitem{Gao:2019twq}
X.~Gao and Z.-B. Yao, \emph{{Spatially covariant gravity theories with two
  tensorial degrees of freedom: the formalism}},
  \href{https://doi.org/10.1103/PhysRevD.101.064018}{\emph{Phys. Rev. D}
  {\bfseries 101} (2020) 064018}
  [\href{https://arxiv.org/abs/1910.13995}{{\ttfamily 1910.13995}}].

\bibitem{Yao2021}
Z.-B. Yao, M.~Oliosi, X.~Gao and S.~Mukohyama, \emph{Minimally modified gravity
  with an auxiliary constraint: A hamiltonian construction},
  \href{https://doi.org/10.1103/PhysRevD.103.024032}{\emph{Phys. Rev. D}
  {\bfseries 103} (2021) 024032}
  [\href{https://arxiv.org/abs/2011.00805}{{\ttfamily 2011.00805}}].

\bibitem{Yao:2023aa}
Z.-B. Yao, M.~Oliosi, X.~Gao and S.~Mukohyama, \emph{Minimally modified gravity
  with auxiliary constraints formalism},
  \href{https://doi.org/10.1103/PhysRevD.107.104052}{\emph{YITP-23-09,
  IPMU23-0003} (2023) } [\href{https://arxiv.org/abs/2302.02090}{{\ttfamily
  2302.02090}}].

\bibitem{Wang:2024hfd}
Z.-C. Wang and X.~Gao, \emph{{Spatial covariant gravity with two degrees of
  freedom in the presence of an auxiliary scalar field: Perturbation
  analysis}}, \href{https://doi.org/10.1088/1674-1137/ad47a9}{\emph{Chin. Phys.
  C} {\bfseries 48} (2024) 085101}
  [\href{https://arxiv.org/abs/2403.15355}{{\ttfamily 2403.15355}}].

\bibitem{Yu:2026fgn}
Y.~Yu, Y.-M. Hu and X.~Gao, \emph{{Spatially covariant gravity with two degrees
  of freedom: A perturbative analysis up to cubic order}},
  \href{https://arxiv.org/abs/2604.14490}{{\ttfamily 2604.14490}}.

\bibitem{Arora:2025msq}
S.~Arora, A.~De~Felice and S.~Mukohyama, \emph{{Dynamical dark energy
  parametrizations in VCDM}},
  \href{https://doi.org/10.1103/l5bx-snl3}{\emph{Phys. Rev. D} {\bfseries 112}
  (2025) 123518} [\href{https://arxiv.org/abs/2508.03784}{{\ttfamily
  2508.03784}}].

\bibitem{Scherer:2025aa}
M.~Scherer, M.~A. Sabogal, R.~C. Nunes and A.~D. Felice, \emph{Challenging
  $\Lambda$CDM: 5$\sigma$ evidence for a dynamical dark energy late-time
  transition},  \href{https://arxiv.org/abs/2504.20664}{{\ttfamily
  2504.20664}}.

\bibitem{Borghetto:2026ytv}
G.~Borghetto, A.~Malhotra, S.~Arora, A.~De~Felice, S.~Mukohyama, G.~Tasinato
  et~al., \emph{{Data-Driven Discovery of a Simple Phantom-Crossing Dark Energy
  Parametrization}},  \href{https://arxiv.org/abs/2606.17951}{{\ttfamily
  2606.17951}}.

\bibitem{Iyonaga2021}
A.~Iyonaga and T.~Kobayashi, \emph{Distinguishing modified gravity with just
  two tensorial degrees of freedom from general relativity: Black holes,
  cosmology, and matter coupling},
  \href{https://doi.org/10.1103/PhysRevD.104.124020}{\emph{Phys. Rev. D}
  {\bfseries 104} (2021) 124020}
  [\href{https://arxiv.org/abs/2109.10615}{{\ttfamily 2109.10615}}].

\bibitem{Hiramatsu2022}
T.~Hiramatsu and T.~Kobayashi, \emph{Testing gravity with the cosmic microwave
  background: constraints on modified gravity with two tensorial degrees of
  freedom}, \href{https://doi.org/10.1088/1475-7516/2022/07/040}{\emph{JCAP}
  {\bfseries 07} (2022) 040}
  [\href{https://arxiv.org/abs/2205.04688}{{\ttfamily 2205.04688}}].

\bibitem{Aoki2020b}
K.~Aoki, A.~De~Felice, S.~Mukohyama, K.~Noui, M.~Oliosi and M.~C. Pookkillath,
  \emph{Minimally modified gravity fitting Planck data better than
  $\Lambda$CDM},
  \href{https://doi.org/10.1140/epjc/s10052-020-8291-1}{\emph{Eur. Phys. J. C}
  {\bfseries 80} (2020) 708}
  [\href{https://arxiv.org/abs/2005.13972}{{\ttfamily 2005.13972}}].

\bibitem{Yao:2025wlx}
Z.~Yao, G.~Ye and A.~Silvestri, \emph{{A general model for dark energy crossing
  the phantom divide}},
  \href{https://doi.org/10.1088/1475-7516/2025/10/078}{\emph{JCAP} {\bfseries
  10} (2025) 078} [\href{https://arxiv.org/abs/2508.01378}{{\ttfamily
  2508.01378}}].

\bibitem{Gao2014}
X.~Gao, \emph{{Unifying framework for scalar-tensor theories of gravity}},
  \href{https://doi.org/10.1103/PhysRevD.90.081501}{\emph{Phys. Rev. D}
  {\bfseries 90} (2014) 081501}
  [\href{https://arxiv.org/abs/1406.0822}{{\ttfamily 1406.0822}}].

\bibitem{Gao2019c}
X.~Gao and Z.-B. Yao, \emph{{Spatially covariant gravity with velocity of the
  lapse function: the Hamiltonian analysis}},
  \href{https://doi.org/10.1088/1475-7516/2019/05/024}{\emph{JCAP} {\bfseries
  05} (2019) 024} [\href{https://arxiv.org/abs/1806.02811}{{\ttfamily
  1806.02811}}].

\bibitem{Langlois:2015cwa}
D.~Langlois and K.~Noui, \emph{{Degenerate higher derivative theories beyond
  Horndeski: evading the Ostrogradski instability}},
  \href{https://doi.org/10.1088/1475-7516/2016/02/034}{\emph{JCAP} {\bfseries
  1602} (2016) 034} [\href{https://arxiv.org/abs/1510.06930}{{\ttfamily
  1510.06930}}].

\bibitem{DeFelice2018}
A.~De~Felice, D.~Langlois, S.~Mukohyama, K.~Noui and A.~Wang,
  \emph{{Generalized instantaneous modes in higher-order scalar-tensor
  theories}}, \href{https://doi.org/10.1103/PhysRevD.98.084024}{\emph{Phys.
  Rev. D} {\bfseries 98} (2018) 084024}
  [\href{https://arxiv.org/abs/1803.06241}{{\ttfamily 1803.06241}}].

\bibitem{Gubitosi:2012hu}
G.~Gubitosi, F.~Piazza and F.~Vernizzi, \emph{{The Effective Field Theory of
  Dark Energy}},
  \href{https://doi.org/10.1088/1475-7516/2013/02/032}{\emph{JCAP} {\bfseries
  1302} (2013) 032} [\href{https://arxiv.org/abs/1210.0201}{{\ttfamily
  1210.0201}}].

\bibitem{Gleyzes:2014rba}
J.~Gleyzes, D.~Langlois and F.~Vernizzi, \emph{{A unifying description of dark
  energy}}, \href{https://doi.org/10.1142/S021827181443010X}{\emph{Int. J. Mod.
  Phys.} {\bfseries D23} (2015) 1443010}
  [\href{https://arxiv.org/abs/1411.3712}{{\ttfamily 1411.3712}}].

\bibitem{Tsujikawa:2007gd}
S.~Tsujikawa, \emph{{Matter density perturbations and effective gravitational
  constant in modified gravity models of dark energy}},
  \href{https://doi.org/10.1103/PhysRevD.76.023514}{\emph{Phys. Rev. D}
  {\bfseries 76} (2007) 023514}
  [\href{https://arxiv.org/abs/0705.1032}{{\ttfamily 0705.1032}}].

\bibitem{Daniel:2008et}
S.~F. Daniel, R.~R. Caldwell, A.~Cooray and A.~Melchiorri, \emph{{Large Scale
  Structure as a Probe of Gravitational Slip}},
  \href{https://doi.org/10.1103/PhysRevD.77.103513}{\emph{Phys. Rev. D}
  {\bfseries 77} (2008) 103513}
  [\href{https://arxiv.org/abs/0802.1068}{{\ttfamily 0802.1068}}].

\bibitem{Frusciante:2016aa}
N.~Frusciante, G.~Papadomanolakis and A.~Silvestri, \emph{An extended action
  for the effective field theory of dark energy: a stability analysis and a
  complete guide to the mapping at the basis of EFTCAMB},
  \href{https://doi.org/10.1088/1475-7516/2016/07/018}{\emph{JCAP}
  {\bfseries 07} (2016) 018}
  [\href{https://arxiv.org/abs/1601.04064}{{\ttfamily 1601.04064}}].

\bibitem{Langlois2017a}
D.~Langlois, M.~Mancarella, K.~Noui and F.~Vernizzi, \emph{{Effective
  Description of Higher-Order Scalar-Tensor Theories}},
  \href{https://doi.org/10.1088/1475-7516/2017/05/033}{\emph{JCAP} {\bfseries
  05} (2017) 033} [\href{https://arxiv.org/abs/1703.03797}{{\ttfamily
  1703.03797}}].

\bibitem{Gleyzes2013}
J.~Gleyzes, D.~Langlois, F.~Piazza and F.~Vernizzi, \emph{{Essential Building
  Blocks of Dark Energy}},
  \href{https://doi.org/10.1088/1475-7516/2013/08/025}{\emph{JCAP} {\bfseries
  08} (2013) 025} [\href{https://arxiv.org/abs/1304.4840}{{\ttfamily
  1304.4840}}].

\bibitem{Frusciante2020}
N.~Frusciante and L.~Perenon, \emph{Effective field theory of dark energy: A
  review}, \href{https://doi.org/10.1016/j.physrep.2020.02.004}{\emph{Phys.
  Rept.} {\bfseries 857} (2020) 1}
  [\href{https://arxiv.org/abs/1907.03150}{{\ttfamily 1907.03150}}].

\bibitem{Horava:2009uw}
P.~Horava, \emph{{Quantum Gravity at a Lifshitz Point}},
  \href{https://doi.org/10.1103/PhysRevD.79.084008}{\emph{Phys. Rev.}
  {\bfseries D79} (2009) 084008}
  [\href{https://arxiv.org/abs/0901.3775}{{\ttfamily 0901.3775}}].

\bibitem{Saito:2024aa}
J.~Saito, Z.~Yao and T.~Kobayashi, \emph{PPN meets EFT of dark energy:
  Post-Newtonian approximation in higher-order scalar-tensor theories},
  {\emph{RUP-24-3} (2024) } [\href{https://arxiv.org/abs/2402.10459}{{\ttfamily
  2402.10459}}].

\bibitem{Schutz:1977df}
B.~F. Schutz and R.~Sorkin, \emph{{Variational aspects of relativistic field
  theories, with application to perfect fluids}},
  \href{https://doi.org/10.1016/0003-4916(77)90200-7}{\emph{Annals Phys.}
  {\bfseries 107} (1977) 1}.

\bibitem{Brown:1993aa}
D.~Brown, \emph{Action functionals for relativistic perfect fluids},
  \href{https://doi.org/10.1088/0264-9381/10/8/017}{\emph{Class. Quant. Grav.}
  {\bfseries 10} (1993) 1579}
  [\href{https://arxiv.org/abs/gr-qc/9304026}{{\ttfamily gr-qc/9304026}}].

\bibitem{Felice:2010aa}
A.~D. Felice, J.-M. Gerard and T.~Suyama, \emph{Cosmological perturbations of a
  perfect fluid and noncommutative variables},
  \href{https://doi.org/10.1103/PhysRevD.81.063527}{\emph{Phys. Rev. D}
  {\bfseries 81} (2010) 063527}
  [\href{https://arxiv.org/abs/0908.3439}{{\ttfamily 0908.3439}}].

\bibitem{Felice:2016aa}
A.~D. Felice, N.~Frusciante and G.~Papadomanolakis, \emph{On the stability
  conditions for theories of modified gravity in presence of matter fields},
  \href{https://doi.org/10.1088/1475-7516/2017/03/027}{\emph{YITP-16-102}
  (2016) } [\href{https://arxiv.org/abs/1609.03599}{{\ttfamily 1609.03599}}].

\bibitem{Aoki:2025aa}
K.~Aoki, J.~B. Jim{\'e}nez, M.~C. Pookkillath and S.~Tsujikawa, \emph{Effective
  field theory of coupled dark energy and dark matter}, {\emph{YITP-25-56,
  WUCG-25-04} (2025) } [\href{https://arxiv.org/abs/2504.17293}{{\ttfamily
  2504.17293}}].

\bibitem{Hell:2025lgn}
A.~Hell and M.~Sasaki, \emph{{Accelerating Universe from constraints}},
  \href{https://doi.org/10.1088/1475-7516/2026/07/034}{\emph{JCAP} {\bfseries
  07} (2026) 034} [\href{https://arxiv.org/abs/2507.00986}{{\ttfamily
  2507.00986}}].

\bibitem{Sawicki2015}
I.~Sawicki and E.~Bellini, \emph{{Limits of quasistatic approximation in
  modified-gravity cosmologies}},
  \href{https://doi.org/10.1103/PhysRevD.92.084061}{\emph{Phys. Rev. D}
  {\bfseries 92} (2015) 084061}
  [\href{https://arxiv.org/abs/1503.06831}{{\ttfamily 1503.06831}}].

\bibitem{Pogosian:2016pwr}
L.~Pogosian and A.~Silvestri, \emph{{What can cosmology tell us about gravity?
  Constraining Horndeski gravity with $\Sigma$ and $\mu$}},
  \href{https://doi.org/10.1103/PhysRevD.94.104014}{\emph{Phys. Rev. D}
  {\bfseries 94} (2016) 104014}
  [\href{https://arxiv.org/abs/1606.05339}{{\ttfamily 1606.05339}}].

\bibitem{Hu:2014aa}
B.~Hu, M.~Raveri, N.~Frusciante and A.~Silvestri, \emph{Effective field theory
  of cosmic acceleration: an implementation in CAMB},
  \href{https://doi.org/10.1103/PhysRevD.89.103530}{\emph{Phys.
  Rev. D} {\bfseries 89} (2014) 103530}
  [\href{https://arxiv.org/abs/1312.5742}{{\ttfamily 1312.5742}}].

\bibitem{Raveri:2014aa}
M.~Raveri, B.~Hu, N.~Frusciante and A.~Silvestri, \emph{Effective field theory
  of cosmic acceleration: constraining dark energy with CMB data},
  \href{https://doi.org/10.1103/PhysRevD.90.043513}{\emph{Phys.
  Rev. D} {\bfseries 90} (2014) 043513}
  [\href{https://arxiv.org/abs/1405.1022}{{\ttfamily 1405.1022}}].

\bibitem{Ye:2026qqf}
G.~Ye, S.~Lin, J.~Pan, D.~de~Boe, S.~Verhoeve, M.~Raveri et~al.,
  \emph{{{\ensuremath{\mathscr{H}}}-EFTCAMB: a Cobaya-integrated,
  Python-wrapped extension of EFTCAMB for covariant Horndeski gravity}},
  \href{https://doi.org/10.1088/1475-7516/2026/07/041}{\emph{JCAP} {\bfseries
  07} (2026) 041} [\href{https://arxiv.org/abs/2603.01662}{{\ttfamily
  2603.01662}}].

\bibitem{Cai:2026swf}
R.-G. Cai and S.-J. Wang, \emph{{The Hubble Tension: A Decade Review}},
  \href{https://doi.org/10.1088/1674-4527/ae842f}{\emph{Res. Astron.
  Astrophys.} {\bfseries 26} (2026) 084011}
  [\href{https://arxiv.org/abs/2606.20434}{{\ttfamily 2606.20434}}].

\bibitem{Pantos:2026koc}
I.~Pantos and L.~Perivolaropoulos, \emph{{Status of the $S_8$ Tension: A 2026
  Review of Probe Discrepancies}},
  \href{https://doi.org/10.1016/j.dark.2026.102286}{\emph{Phys. Dark Univ.}
  {\bfseries 52} (2026) 102286}
  [\href{https://arxiv.org/abs/2602.12238}{{\ttfamily 2602.12238}}].

\end{thebibliography}
\end{document}